\documentclass[article,amsmath,11pt,amssymb,floatfix,superscriptaddress,
showpagenumber, nofootinbib]{revtex4}
\usepackage{graphicx}
\usepackage{epsfig}
\usepackage{bm}
\usepackage{amsfonts}
\usepackage[mathscr]{euscript}
\usepackage{xcolor}
\usepackage{caption,subcaption}
\input epsf

\begin{document}
\title{\Large Thermodynamics of the most generalized form of Holographic Dark Energy and some particular cases with Corrected Entropies }
\author{Sanghati Saha}
\email{sanghati.saha1504@gmail.com; sanghati.saha2@s.amity.edu}
\affiliation{ Department of Mathematics, Amity University, Kolkata, Major
Arterial Road, Action Area II, Rajarhat, New Town, Kolkata 700135,
India.}
\author{Ertan Güdekli}
\email{gudekli@istanbul.edu.tr}
\affiliation{ Department of Physics, Istanbul University, Istanbul 34134, Turkey.}
\author{Surajit Chattopadhyay}
\email{schattopadhyay1@kol.amity.edu; surajitchatto@outlook.com}
\affiliation{ Department of Mathematics, Amity University, Kolkata, Major
Arterial Road, Action Area II, Rajarhat, New Town, Kolkata 700135, India.(Communicating author)}
\date{\today}
\begin{abstract}
The holographic cut-off in generalized dark energy (HDE) formalism depends on its cut-off. In accordance with this, a four-parameter generalized entropy has recently been developed. For appropriate limits of the parameters, it reduces to various known entropies in the study of Odintsov, S. D., S. D’Onofrio, and T. Paul. (2023) \textit{Physics of the Dark Universe}, \textbf{42} pp: 101277. In the current work, we investigate the evolution of the universe in its early phase and late phase within the framework of entropic cosmology, where the entropic energy density functions are reconstructed within the framework of the equivalence of holographic dark energy and four-parameter generalized entropy ($\mathcal{S}_g$). Along with the aforementioned reconstruction scheme, in this study, we demonstrate that an extensive variety of dark energy (DE) models can be considered as distinct and particular candidates for the most generalized four-parameter entropic HDE family, each having their own cut-off. We examined a number of entropic dark energy models in this regard, including the generalized holographic dark energy with Nojiri-Odintsov(NO) cut-off, the Barrow entropic HDE (BHDE) with particle horizon as IR cut-off, the Tsallis entropic HDE (THDE) with future event horizon as IR cut-off, all of three cases are particular cases of the most generalized four parameter entropic holographic dark energy. Inspired by S. Nojiri, and S. D. Odintsov (2006) (\textit{General Relativity and Gravitation}, \textbf{38} p: 1285-1304 )  and (S. Nojiri and S. D. Odintsov, 2017, \textit{European Physical Journal C}, \textbf{77}, pp.1-8 ); our current work reports a study on cosmological parameters and thermodynamics with entropy-corrections (logarithmic and power-law) to cosmological horizon entropy as well as black hole entropy with a highly generalized viscous coupled holographic dark fluid along its particular cases. 
\end{abstract}
 
\maketitle
\section{Introduction}
The current phase of our Universe's expansion is accelerated, according to numerous cosmological evaluations \cite{acceleration1,acceleration2,acceleration3,acceleration4,acceleration5}. A plausible explanation for this cosmic acceleration could involve examining exotic matter, known as dark energy (DE), which accounts for roughly $68\%$ of the universe's entire energy budget. The type and source of this DE, however, remain completely unknown. Conversely, dark matter (DM), which makes up around $28\%$ of the universe's total energy density, is the second-largest component. Similar to the DE sector, little is known about the DM sector. To account for the current acceleration phase, several theoretical models have been considered up to this point; some outstanding reviews on this topic may be found in \cite{acceleration6, acceleration7, acceleration8}. However, the origin and nature of cosmic acceleration continue to be a mystery in contemporary cosmology. In this context, holographic dark energy (HDE) is an intriguing attempt to address this issue (see to \cite{HDE1,HDE2,HDE3} for specifics), some of which are outlined in \cite{HDE4,HDE5,HDE6,HDE7,HDE8,HDE NO,HDE10,HDE11,HDE12,HDE13,HDE14,HDE15,HDE16,HDE17,HDE18,HDE19,HDE20,HDE21,HDE22,HDE23,HDE24,HDE25,HDE26,HDE27,HDE28}. Based on the principles of black hole thermodynamics and string theory, the holographic principle links the maximum distance of a quantum field theory to its infrared cutoff, which is associated with the vacuum energy. In the field cosmology, the holographic approach is often used, especially when characterizing the dark energy (DE) era, which is commonly referred to as the holographic dark energy (HDE) model \cite{HDE4,HDE5,HDE6,HDE7,HDE8,HDE NO,HDE10,HDE11,HDE12,HDE13,HDE14,HDE15,
HDE16,HDE17,HDE18,HDE19,HDE20,HDE21,HDE22,HDE23,HDE24,HDE25,HDE26,HDE27,HDE28}. It is important to emphasize that the HDE differs greatly from the other dark energy theories since it is based on the holographic principle and dimensional analysis instead of adding an additional term in the Lagrangian. The holographic concept has been successfully applied to the early inflationary Universe in addition to the dark energy era \cite{HDEinflation1, HDEinflation2, HDEinflation3, HDEinflation4}. Since the Universe was actually quite small in its early stages, the holographic energy density played a crucial role in starting an inflationary scenario. In addition, the holographic inflation was shown to be consistent with the 2018 Planck constraints. From an alternative perspective, the holographic principle was applied to the bouncing scenario in \cite{HDEbounce1, HDEbounce2, HDEbounce3}, where the holographic energy density contributes to the violation of the energy conditions, which causes the universe to bounce. From an alternative perspective, the holographic principle was applied to the bouncing scenario in [44–48], where the holographic energy density contributes to the violation of the energy conditions, which causes the universe to bounce. Resuming the discussion of dark energy, we find that the holographic dark energy density is proportional to the inverse squared of the holographic cut-off, commonly considered the particle horizon or the future horizon. But in this case, there is still disagreement over the basic structure of the holographic cut-off. The most generalized cut-off has been proposed in [9], where the cut-off is specifically thought to depend on the future event horizon, particle horizon, and their time derivatives. This leads to the generalized version of HDE, or \textquotedblleft{generalized HDE \textquotedblright}. In this study we have chosen generalized holographic dark fluid with Nojiri-Odintsov(NO) cut-off influenced by \cite{HDE NO}. The feasibility of unifying the early- and late-time universes based on phantom cosmology is illustrated in \cite{HDE NO}, where a thorough examination of the reason behind considering such cut-offs is found. The notion of phantom non-phantom transition, which develops so that the universe could have effectively phantom equation of state at early and late time, is also one of the intriguing conclusions (among the others) described in \cite{HDE NO}. In general, the oscillating universe could have some phantom and non-phantom phases. Again the Tsallis holographic dark energy (THDE) is a new HDE model that has been suggested specifically utilizing the Tsallis entropy and the holographic hypothesis \cite{HDE TSallis1,HDE TSallis2,HDE TSallis3} . Consequently,  THDE models have been studied and examined in many contexts recently to find the universe's dynamics; for a thorough overview, see \cite{HDE TSallis1,HDE TSallis2,HDE TSallis3,HDE TSallis4,HDE TSallis5,HDE TSallis6,HDE TSallis7,HDE TSallis8,HDE TSallis9}. Conversely, the Barrow entropic dark energy model \cite{Barrow1,Barrow2,Barrow3,Barrow4,Barrow5,Barrow6,Barrow7} has garnered significant interest recently since it provides a reasonable explanation for the universe's dark energy era. It was demonstrated in \cite{Barrow1} that quantum gravitational effects might add fractal patterns on black hole formations, which could be recorded within the entropy function $S \propto A^{(1+\Delta)}$, where A stands for the standard horizon area. This led to the proposal of the Barrow entropy. The range of the exponent $\Delta$ is $0 < \Delta<1$, and when $\Delta = 0$, it matches the standard Bekenstein-Hawking entropy function. During the cosmological evolution, the Barrow entropic dark energy equation of state parameter may lead to a quintessence region, phantom region, or phantom crossover, depending on the value of $\Delta$. Cosmologists are increasingly interested in various entropies, such as the Tsallis entropy, the Renyi entropy, the Barrow entropy, the Sharma-Mittal entropy, the Kaniadakis entropy, and the Loop Quantum Gravity Entropy, leading to the development of a most generalized entropy. Keeping this in mind, researchers have developed a generalized entropy that demonstrates the reduction of all known entropies with an appropriate limit of the entropic parameters \cite{four parameter1, four parameter2, four parameter3, four parameter4, four parameter5}. Two different generalized entropies have been suggested, one with six parameters and the other with four. However, a theory published in \cite{four parameter2} said that a generalized entropy function only needs four parameters to include all other entropies. In the context of bounce cosmology, both the generalized entropies (with 6 parameters and 4 parameters) exhibit a divergent nature at the instant of $H = 0$ (where H is the Hubble parameter). In this regard, we proposed a conjecture stating, \textquotedblleft{the minimum number of parameters required in a generalized entropy function that can generalize all the known entropies and is also singular-free during the universe's evolution is equal to five.\textquotedblright} In the present paper, we intend to consider the four parameters of generalized entropy. We will examine its possible thermodynamic implications for the holographic reconstruction of this most generalized entropy during the early and late-time universe. The recently suggested generalized entropy unifies various recognized and seemingly disparate entropies, such as the Tsallis, Renyi, Barrow, Kaniadakis, Sharma-Mittal, and loop quantum gravity entropies, under a single heading. the microscopic interpretation of horizon entropy is still a debatable topic, however one may see \cite{microscope}. A microscopic thermodynamic explanation of generalized entropy(ies) from canonical and grand-canonical ensembles is given by the authors of \cite{microscope}.

Viscosity coefficients have long been included in cosmology, but it has long been believed that the physical significance of these phenomenological factors is negligible or, at the very least, subdominant. The most significant influence of viscosity on the very early universe is believed to have occurred when the temperature reached about $10^{10}$ K during the neutrino decoupling (end of the lepton epoch). Viscosity was first introduced from a particle physics perspective by Misner \cite{Viscosity1}; see also Zel'dovich and Novikov \cite{Viscosity2}. On the other hand, Eckart's work was the first to present the viscosity concept from a phenomenological standpoint \cite{bulk1}. For several reasons in cosmology, viscosity theories have gained increased attention in recent years, perhaps most notably from an elementary standpoint. The brane-bulk energy exchange term in viscous brane cosmology was shown in \cite{bulk2}. They were particularly interested in the viscous fluid on the brane's energy conservation equation. In contrast to the proper equation derived from the Boltzmann equation, it turned out that the emission process equates to a negative entropy change for the thermodynamic subsystem on the brane. Using a formula for the entropy of a multicomponent correlated fluid, the authors of \cite{bulk3} established an association between the entropy of a closed FRW universe and its energy, including Casimir energy. Under some conditions, this formula reduces to the Cardy-Verlinde form. An inhomogeneous equation of state was adhered to by the generalized viscous fluid.  The famous Cardy-Verlinde formula corresponding to the $2D$ CFT entropy is demonstrated to be reduced to in some rare cases in such an equation. This leads \cite{bulk4} to propose a viscous Little Rip cosmology. The authors find that whereas bulk viscosity normally favors the Big Rip, there are particular situations when the formalism readily converts to the Little Rip scenario. In particular, they showed that a Little Rip cosmology can arise in a viscous fluid with an inhomogeneous (imperfect) equation of state, or, equivalently, as a pure viscosity effect in one. The combined effect of viscosity and a generic (power-like) equation of state is also thoroughly investigated. \cite{bulk5} reports on a newly developed viscous Little Rip cosmology in an isotropic fluid. Their studies focused on characterizing the turbulent fluid in the late universe. They examined the possible evolution of the Universe from the one-component example of a viscous era with constant bulk viscosity to a turbulent era. Using the decay of isotropic turbulence in ordinary hydrodynamics, in FRW universe, the effects of an initial fraction $f$ of turbulent kinetic energy in the cosmic fluid on the cosmological development in the late quintessence/phantom universe are investigated in \cite{bulk6}. 

The field is usually assumed to be non-interacting in dark energy models. Presently , densities of dark energy and dark matter are equivalent, and by assuming the right interaction between the two dark sectors, it is easy to solve the coincidence problem. The transition from matter domination to dark energy domination can be elucidated by an adequate energy exchange rate. In order to get a satisfactory evolution of the universe, it is commonly assumed that there is an interaction term \cite{interaction1,interaction2,interaction3}. This interaction term is based on the idea that the decay rate is proportional to the current value of the Hubble parameter, which allows for a good match to the universe's expansion history. However, several alternative forms of interactions have also been contemplated, in which the interaction term has been selected based on phenomenological considerations. So, motivated by those aforementioned facts of viscosity and interaction, we have assumed viscous interacting scenario of different faces of HDE models.

The relationship between thermodynamics and Einstein field equations was initially observed in \cite{thermodynamic1} within the framework of Einstein gravity. This was achieved by deducing the Einstein equation from the correlation between entropy and horizon area and applying the first law of thermodynamics, $\delta Q$, in the Rindler spacetime. According to black hole thermodynamics, a black hole's temperature and horizon area are related to its entropy and surface gravity, respectively \cite{thermodynamic2,thermodynamic3}. A Friedmann equation similar to the Cardy-Verlinde formula, an entropy formula for a conformal field theory, can be expressed in a radiation-dominated Friedmann–Robertson–Walker (FRW) Universe \cite{thermodynamic4}. It is commonly known that event horizons, whether they are cosmological or black holes, resemble black bodies and have non-vanishing entropy and temperature, with the latter following the Bekenstein–Hawking entropy formula \cite{thermodynamic5,thermodynamic6}. The equation $\mathcal{S} = \frac{A}{4}$ represents the relationship between the surface area $\mathcal{S}$ and the area of the horizon $(A)$ with $(c=\mathscr{G}=h=1)$, where, $A$ is defined as $4\pi R_h^2$, where $R$ is the radius of the horizon, $h$ is the area of the horizon,and $\mathscr{G}$ is Newton’s gravitational constant. The equation $-dE = \mathcal{T} d\mathcal{S}$ represents the first law of thermodynamics for the cosmic horizon andin this equation, $\mathcal{T} = \frac{1}{2 \pi R_h}$ represents the Hawking temperature \cite{thermodynamic7,thermodynamic8}. It was recently shown that cosmological apparent horizons also have thermodynamical characteristics, which are formal equivalent to those of event horizons \cite{thermodynamic9}. In a de Sitter space-time with flat spatial geometry, the event horizon and the apparent horizon of the Universe are the same, resulting in a single cosmological horizon. It was discovered that the first law and generalized second law (GSL) of thermodynamics hold on the apparent horizon when the apparent horizon and the event horizon of the universe differ, but they break down if one considers the event horizon \cite{thermodynamic10}. It has recently been shown that black holes will lose mass and finally vanish completely if phantom energy dominates the expansion of the universe \cite{thermodynamic11,thermodynamic12,thermodynamic13,thermodynamic14,thermodynamic15,thermodynamic16,thermodynamic17}. Since these collapsed items are the universe's most entropic creatures, this poses a threat to the GSL \cite{thermodynamic11}. Based on this brief analysis, the researchers are motivated to investigate the thermodynamic implications of Universes dominated by phantom energy. The fact that ever-accelerating universes have a future event horizon, sometimes known as a cosmological horizon, must be considered when doing this \cite{thermodynamic11}. The thermodynamic properties of the apparent horizon have been discovered in a quasi-de Sitter geometry of the inflationary Universe \cite{thermodynamic18}. Setare \cite{thermodynamic19} proposed the use of the interacting holographic model of dark energy to examine the accuracy of the generalized second laws of thermodynamics in a non-flat (closed) Universe that is bounded by the event horizon. The validity of GSL for a system comprising dark energy, dark matter, and radiation in the FRW Universe is demonstrated in \cite{thermodynamic20}. It has also been demonstrated that, in Horava–Lifshitz cosmology, the generalized second law is conditionally valid for an open Universe and generally valid for flat and closed geometry under detailed balance. Still, it is only conditionally valid for all curvatures beyond detailed balance \cite{thermodynamic21}. The GSL has been expanded in a detailed analysis to encompass a range of generalized gravity theories, including Lovelock, Gauss-Bonnet, braneworld, scalar-tensor, and f(R) models \cite{thermodynamic22}. In the context of Einstein's theory of gravity, the entropy of the horizon is directly proportional to the horizon's area, denoted as $\mathcal{S} \propto A$. When the theory of gravity is altered by incorporating additional curvature factors into the action principle, it modifies the entropy-area relation. For example, in f(R) gravity, the relation is expressed as$\mathcal{S} \propto f(R) A$ \cite{thermodynamic23}. However, quantum corrections, including power-law and logarithmic corrections, have been applied to the semi-classical entropy law in recent years. Due to variations in quantum fluctuations and thermal equilibrium, loop quantum gravity results in logarithmic corrections \cite{thermodynamic24,thermodynamic25,thermodynamic26,thermodynamic27,thermodynamic28,thermodynamic29,thermodynamic30} which can be written as;
\begin{equation}
\mathcal{S}_h=\frac{A}{4G}+\upsilon \ln{\frac{A}{4G}}+\Psi \frac{4 G}{A}+\psi 
\label{logarithimic horizon}
\end{equation}
In dealing with the entanglement of quantum fields in and out of the horizon, on the other hand, power-law corrections are observed \cite{thermodynamic31,thermodynamic32,thermodynamic33,thermodynamic34,thermodynamic35};
\begin{equation}
\mathcal{S}=\frac{A}{4} [1-K_\alpha A^{1-\frac{\alpha}{2}}],
\label{powerlaw entropy}
\end{equation}
where, $A=4\pi R_h^2$ and \textbf{$K_\alpha=\frac{\alpha(4\pi)^{\frac{\alpha}{2}-1}}{(4-\alpha)r_c^{2-\alpha}}$}.
In this context, $r_c$ represents the crossover scale, while $\alpha$ is a dimensionless constant that is now a subject of discussion regarding its precise value. The second part in Eq. (\ref{powerlaw entropy}) can be interpreted as a power-law correction to the entropy-area relationship. This modification arises from the entanglement of the scalar field's wave function between the ground and excited states, as discussed in references \cite{thermodynamic31,thermodynamic32,thermodynamic33,thermodynamic34,thermodynamic35}. Furthermore, with larger excitations, the correction term has greater significance. Acknowledging that the correction term decreases more rapidly with A is crucial. Therefore, the entropy-area law is restored in the semi-classical limit  (large area). 

The structure of the paper is outlined as follows: Reconstruction schemes for different holographic fluids are presented in Section II, where a bulk viscous framework is considered. The generalized second law of thermodynamics with Bekenstein entropy of HDE with different cut-offs is presented in Section III. Thermodynamics under corrected entropies is presented in Section IV. We have concluded in Section V.

\section{RECONSTRUCTION SCHEME FOR DIFFERENT FACES OF HOLOGRAPHIC DARK FLUID IN VISCOUS INTERACTING SCENARIO}
The FRW metric for a homogeneous and isotropic flat universe is given by,
\begin{equation}
ds^2=-dt^2+a^2(t)(dr^2+r^2(d\theta^2+sin^2\theta d\phi^2),
\label{FRWmetric}
\end{equation}
in which $t$ represents the cosmic time and $a(t)$ is the scale factor. Here, we have defined the units such that $8 \pi G = 1$. In the context of isotropic and homogeneous cosmologies, every dissipation process in a FRW cosmology can be described as scalar. One well-known consequence of the FRW cosmic solutions, which correspond to universes with perfect fluid and bulk viscous stresses, is the possibility of DEC violations \cite{dec1,dec2}. In Eckart's theory, the bulk viscous pressure $\Pi$ (referenced in \cite{bulk1}) can be expressed as $\Pi=-3H\xi$, where $\xi$ represents the fluid's bulk viscosity and is determined by the Hubble parameter and transport coefficient. In \cite{xi1}, $\xi \propto \rho$ and $\rho \propto H$ are seen. In their study, Meng and Dou \cite{xi2} propose a relationship between the bulk viscosity coefficient and the Hubble parameter H. In this study, we have considered the function $\xi(H)=\xi H$, where $\xi$ is a constant parameter, based on the works described above. Hence, the resultant model reads the bulk viscous pressure of the universe as $\Pi=-3\xi H^2$. Due to the importance of bulk viscosity \cite{viscosity3, viscosity4, viscosity5, viscosity6, viscosity7, viscosity8, viscosity9, viscosity10, viscosity11, viscosity12, viscosity13, viscosity14, viscosity15, viscosity16, viscosity17}, we have chosen both the different faces of a universe filled with holographic dark energy (HDE) and dark matter (DM) without taking into account the impact of baryonic matter in this case. The two Friedmann equations can be mathematically represented as:
\begin{equation}
H^2=\frac{1}{3}\rho_{effective}
\label{1st-Fried}
\end{equation}
and, \begin{equation}
6 \frac{\ddot{a}}{a}=-(\rho_{effective}+3(P_{effective}+\Pi))
\label{2nd-Fried}
\end{equation}
according to which, $\rho_{effective}=\rho_{DM}+\rho_{HDE}$. In this context, we are examining pressure-less dark matter, characterized by its density denoted as $\rho_{DM}$ and having a pressure of $P_{DM}=0$. The effective pressure, denoted as $P_{effective}$, is simply redefined by the bulk viscosity, which leads to dissipation. It can be expressed as $P_{effective}=P_{HDE}+\Pi= P_{HDE}-3\xi H^2$.  Again, The following represents the continuity equation:
\begin{equation}
\dot{\rho}_{effective}+3 H (\rho_{effective}+P_{effective}) = 0.
\label{continuity-equation1}
\end{equation}
In the interaction scenario, the continuity equation is reformulated into distinct equations that represent the components of the universe being studied. We have $\rho_{effective}=\rho_{HDE}+\rho_{DM}$ and $P_{Total}=P_{HDE}+\Pi$ in this instance. The following can therefore be written: 
\begin{equation} 
\dot{\rho}_{HDE}+\dot{\rho}_{DM}+3 H (\rho_{HDE}+\rho_{DM}+P_{HDE}+\Pi) = 0. 
\label{continuity-equation2} 
\end{equation}
To meet the equation (\ref{continuity-equation1}), the interaction scenario involves rephrasing the above equation by taking $Q$ in one equation and $-Q$ in the other to neutralize each other. Consequently, using the equation of continuity, we obtain: 
\begin{equation} 
\dot{\rho}_{DM}+3 H \rho_{DM} = Q,
\label{conservation1} 
\end{equation} 
\begin{equation} 
\dot{\rho}_{HDE}+3 H (\rho_{HDE}+P_{HDE}+\Pi)  = -Q.
\label{conservation2}
\end{equation} 
In the above equations, $Q$ \cite{interaction1,interaction2,interaction3} represents the interaction term between HDE and DM. Based on the literature \cite{interaction1}, we have assumed $Q=3Hb^2\rho_{DM}$ \cite{interaction Q} in the current work. In order to analyze the reconstructed scenario, we will now give a brief overview of the key characteristics of the three scale factors that were considered—the emergent, logamediate, and intermediate ones—and derive some relevant physical quantities from them.
\subsection{Cosmological Settings of Four Parameter Entropic Most Generalized Holographic Dark fluid}
Here we will briefly discuss the generalized four-parameter entropy function, first proposed in \cite{four parameter2}, which at certain parameter limits can reproduce the known entropies described thus far. Let us begin with the Bekenstein-Hawking entropy \cite{thermodynamic2, thermodynamic3} $\mathcal{S}_h=\frac{A}{4G}$ in this regard, where the area of the horizon is represented by $A=4\pi R_h^2$ and the horizon radius is denoted by $R_h$. Thus, various entropy functions—the Tsallis entropy\cite{HDE TSallis1, HDE TSallis2} , the Renyi entropy\cite{Renyi} , the Barrow entropy\cite{Barrow1,Barrow2}, the Sharma-Mittal entropy\cite{Sharma Mittal}, the Kaniadakis entropy\cite{Kanidakis}, and the Loop Quantum gravity entropy\cite{Loop Quantum}—have been introduced based on the system under exploration. Naturally, the growing interest in these various entropies in cosmology and black hole physics begs the question, "Is there a unique or generalized entropy that can generalize all these known entropies?" In keeping with this, two distinct generalized entropy functions with six and four parameters, respectively, were proposed by a few of our authors\cite{four parameter1,four parameter2,four parameter3,four parameter4,four parameter5}. These functions can generalize all of the previously described known entropies that have been proposed thus far. Specifically, the generalized entropies can be obtained from,
\begin{equation}
\text{6 parameter entropy:}~
\mathcal{S}_6[\alpha_{\pm},\beta_{\pm},\gamma_{\pm}]=\frac{1}{\alpha_{+}+\alpha_{-}}[(1+\frac{\alpha_{+}}{\beta_{+}}\mathcal{S}^{\gamma_{+}})^{\beta_{+}}-(1+\frac{\alpha_{-}}{\beta_{-}}\mathcal{S}^{\gamma_{-}})^{-\beta_{-}}]
\label{6-parameter entropy}
\end{equation}
\begin{equation}
\text{4 parameter entropy:}~
\mathcal{S}_g[\alpha_{+},\alpha_{-},\beta,\gamma]=\frac{1}{\gamma}[(1+\frac{\alpha_{+}}{\beta}\mathcal{S})^{\beta}-(1+\frac{\alpha_{-}}{\beta}\mathcal{S})^{-\beta}]
\label{4-parameter entropy}
\end{equation}
The entropic parameters of the corresponding generalized entropy, which are taken to be positive. All known entropies that have been proposed thus far (such as the Tsallis entropy, the Renyi entropy, the Barrow entropy, the Sharma-Mittal entropy, the Kaniadakis entropy, and the Loop Quantum gravity entropy) have been explicitly shown to be given by either the 4- or 6-parameter generalized entropy in certain limits\cite{four parameter1,four parameter2,four parameter3,four parameter4,four parameter5}. The 4-parameter generalized entropy function needs a minimum of four parameters in order to generalize all of the entropies that were previously mentioned. Because of this simple construction, the 4-parameter generalized entropy will be examined in this paper. Moreover, the 4-parameter generalized entropy function in Eq. (\ref{4-parameter entropy}) has the same characteristics as the one below: 
(1) For $\mathcal{S}\to 0$, $\mathcal{S}_g \to 0$.\\
(2) With $\mathcal{S}$, the entropy $\mathcal{S}_g[\alpha_{+},\alpha_{-},\beta,\gamma]$ is a monotonically increasing function.The Bekenstein-Hawking entropy appears to be approached by  $\mathcal{S}_g[\alpha_{+},\alpha_{-},\beta,\gamma]$ at a certain limit of the parameters indicated by $\alpha_{+}\to \infty$, $\alpha_{-}=0$,$\gamma=(\alpha_{+}/\beta)^\beta$, and $\beta=1$.

In entropic cosmology, the energy density generated by the entropy function in the Friedmann equations may eliminate the need for an additional scalar field to be manually added in order to characterize the universe's evolution. As a result, and in keeping with the previous discussions, we will now consider the four parameter generalized entropy (represented by $\mathcal{S}_g$; see Eq. (\ref{4-parameter entropy})), which is, in fact, more generalized than the three-parameter entropy function and also more rational than the six-parameter entropy because it has lesser parameters. We will also investigate whether the $\mathcal{S}_g$ can effectively initiate the early stages of the universe. Again, the term \textquotedblleft{holographic dark energy\textquotedblright} refers to a specific concept that can be expressed as follows:
\begin{equation}
\rho_{HDE}=\frac{3\zeta^2}{\mathcal{L}^2}
\label{HDE}
\end{equation}
where, $\zeta$ be the numerical constant and ($\mathcal{L}$) being the infrared cut-off. By choosing infrared cut-off as 4-parameter most generalized entropy and $\mathcal{S}=\frac{A}{4G}~~\text{with}~8\pi G=1$ in this regard, where the area of the horizon is represented by $A=4\pi R_H^2$, we can get simplest form of infrared cut-off:
\begin{equation}
\mathcal{L}=\frac{-\left(1+\frac{\pi \alpha_{-}  R_H^2}{G \beta }\right)^{-\beta }+\left(1+\frac{\pi  \alpha_{+}  R_H^2}{G \beta }\right)^{\beta}}{\gamma},
\label{infrared-four parameter}
\end{equation}
where, $R_H=\frac{1}{H}$ is chosen as Hubble horizon. To find Hubble parameter in this case, we have to choose scale factor in this regard. Here, we have chosen hybrid scale factor \cite{Hybrid Scale factor1,Hybrid Scale factor2,Hybrid Scale factor3,Hybrid Scale factor4} as
\begin{equation}
a= e^{\lambda  t} t^{\delta }
\label{hybrid scale factor}
\end{equation}
where, the constants $\lambda$ and $\delta$ are positive. The hybrid scale factor can explain the cosmic transit from early time acceleration to late time acceleration. Eq. (\ref{hybrid scale factor}) establishes that the exponential component predominates in the late phase of cosmic evolution, while power law behavior dominates the cosmic dynamics in the early phase. The exponential law is restored when $\delta = 0$, while the power law is reduced when $\lambda = 0$. For the hybrid scale factor, the Hubble parameter and Hubble horizon can be derived as;
\begin{equation}
H=\frac{\delta }{t}+\lambda, 
\label{Hubble-hybrid}
\end{equation}
\begin{equation}
R_H=\frac{1}{H}=\frac{t}{\delta+\lambda t}.
\label{R-H}
\end{equation}
Therefore, by choosing $R_H=\frac{1}{H}$ in Eq. (\ref{infrared-four parameter}) and from Eq. (\ref{HDE}), we can get 4-parameter entropic most generalized holographic dark energy(4-parameter GHDE) density;  
\begin{equation}
\rho_{GHDE}=\frac{3 \gamma ^2 \zeta ^2}{\left(-\left(1+\frac{\pi  \alpha_{-} }{G \beta  \left(\frac{\delta }{t}+\lambda \right)^2}\right)^{-\beta
}+\left(1+\frac{\pi  \alpha_{+} }{G \beta  \left(\frac{\delta }{t}+\lambda \right)^2}\right)^{\beta }\right)^2}
\label{4-parameter GHDE density}
\end{equation}
Thus, the expression of dark matter density in this reconstruction approach may be obtained from Eqs. (\ref{Hubble-hybrid}), (\ref{conservation1}), and $Q=3Hb^2\rho_{DM}$;
\begin{equation}
\rho_m=e^{3 \left(-1+b^2\right) (t \lambda +\delta  \log t)} C_1
\label{4-parameter dark-matter density}
\end{equation}
Hence, effective density of reconstructed viscous coupled 4-parameter entropic most generalized holographic dark fluid(4-parameter GHDE) will be:
\begin{equation}
\rho_{effective_{GHDE}}=C_1 e^{3 \left(-1+b^2\right) (t \lambda +\delta  \log t)}+\frac{3 \gamma ^2 \zeta ^2}{\left(\left(1+\frac{\pi
t^2 \alpha_{-} }{G \beta  (\delta +t \lambda )^2}\right)^{-\beta }-\left(1+\frac{\pi  t^2 \alpha_{+} }{G \beta  (\delta +t \lambda )^2}\right)^{\beta }\right)^2}
\label{total density 4-parameter}
\end{equation}
Thus, in this GHDE reconstructed scenario, we can obtain the thermodynamic pressure of 4-parameter GHDE from Eq. (\ref{2nd-Fried}), and the bulk viscous pressure from Eq. (\ref{Hubble-hybrid}) and $\Pi=-3 \xi H^2$. Then, the effective EoS parameter $w_{effective}=\frac{P_{GHDE}+\Pi}{\rho_{effective_{ GHDE}}}$ and the EoS parameter $w=\frac{P_{GHDE}}{\rho_{effective_{GHDE}}}$ can be represented as follows:
\begin{equation}
\begin{array}{c}
P_{GHDE}=-\frac{1}{3} C_1 e^{3 \left(-1+b^2\right) (t \lambda +\delta  \log t)}-\frac{\gamma ^2 \zeta ^2}{\left(\left(1+\frac{\pi  t^2\alpha_{-} }{G \beta  (\delta +t \lambda )^2}\right)^{-\beta }-\left(1+\frac{\pi  t^2 \alpha_{+} }{G \beta  (\delta +t \lambda)^2}\right)^{\beta }\right)^2}\\+\frac{-2
\left((-1+\delta ) \delta +2 t \delta  \lambda +t^2 \lambda ^2\right)+3 (\delta +t \lambda )^2 \xi }{t^2};
\end{array}
\label{pressure 4-parameter}
\end{equation}
\begin{equation}
\begin{array}{c}
w_{GHDE}=\frac{\mathcal{M}_{11}}{9 \gamma ^2 \zeta^2 (\delta +t \lambda )}
\end{array}
\label{w-4-parameter}
\end{equation}
\begin{equation*}
\begin{array}{c}
\mathcal{M}_{11}=t \left(\left(1+\frac{\pi  t^2 \alpha_{-} }{G \beta  (\delta +t \lambda )^2}\right)^{-\beta }-\left(1+\frac{\pi  t^2 \alpha_{+} }{G \beta  (\delta+t \lambda )^2}\right)^{\beta }\right)^2 \\(-3 b^2 C_1 e^{3 \left(-1+b^2\right) (t \lambda +\delta  \log t)} \left(\frac{\delta }{t}+\lambda\right)-\frac{9 \gamma ^2 \zeta ^2 \left(\frac{\delta }{t}+\lambda \right)}{\left(\left(1+\frac{\pi  t^2 \alpha_{-} }{G \beta  (\delta +t \lambda )^2}\right)^{-\beta}-\left(1+\frac{\pi  t^2 \alpha_{+} }{G \beta  (\delta +t \lambda )^2}\right)^{\beta }\right)^2}+\\\frac{12 \pi  \gamma ^2 \delta  \zeta ^2 \left(\alpha_{-} \left(1+\frac{\pi  t^2 \alpha_{-} }{G \beta  (\delta +t \lambda )^2}\right)^{-1-\beta }+\alpha_{+}  \left(1+\frac{\pi  t^2 \alpha_{+} }{G \beta  (\delta +t\lambda )^2}\right)^{-1+\beta }\right)}{G t^2 \left(\frac{\delta }{t}+\lambda \right)^3 \left(-\left(1+\frac{\pi  t^2 \alpha_{-} }{G \beta  (\delta +t\lambda )^2}\right)^{-\beta }+\left(1+\frac{\pi  t^2 \alpha_{+} }{G \beta  (\delta +t \lambda )^2}\right)^{\beta }\right)^3});
\end{array}
\end{equation*}
\begin{equation}
\begin{array}{c}
w_{eff_{GHDE}}=-\frac{\frac{1}{3} C_1 e^{3 \left(-1+b^2\right) (t \lambda +\delta  \log t)}+\frac{2 (-1+\delta ) \delta }{t^2}+\frac{4
\delta  \lambda }{t}+2 \lambda ^2+\frac{\gamma ^2 \zeta ^2}{\left(\left(1+\frac{\pi  t^2 \alpha_{-} }{G \beta  (\delta +t \lambda )^2}\right)^{-\beta
}-\left(1+\frac{\pi  t^2 \alpha_{+} }{G \beta  (\delta +t \lambda )^2}\right)^{\beta }\right)^2}}{C_1 e^{3 \left(-1+b^2\right) (t \lambda +\delta
 \log t)}+\frac{3 \gamma ^2 \zeta ^2}{\left(\left(1+\frac{\pi  t^2 \alpha_{-} }{G \beta  (\delta +t \lambda )^2}\right)^{-\beta }-\left(1+\frac{\pi
 t^2 \alpha_{+} }{G \beta  (\delta +t \lambda )^2}\right)^{\beta }\right)^2}}.
\end{array}
\label{weff-4-parameter}
\end{equation}
\begin{figure}
\centering
\begin{minipage}{.5\textwidth}
  \centering
\includegraphics[width=.9\linewidth]{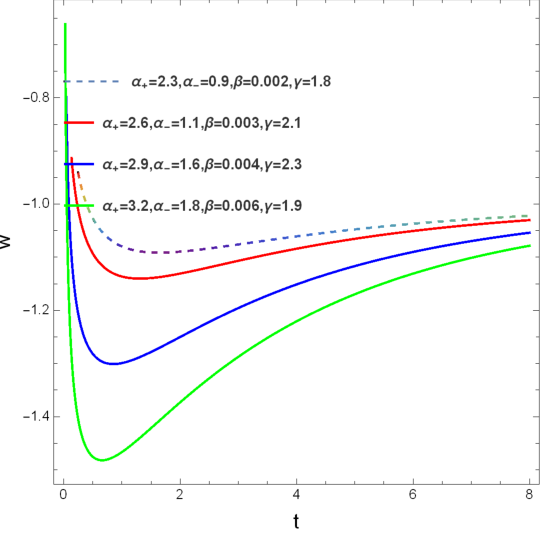}
\caption{Evolution of reconstructed viscous interacting 4-parameter entropic GHDE EoS parameter for the range $b~\text{lies in}~(0,1)$ in hybrid-scale factor scenario.}
\label{Fig_01_GHDE_w}
\end{minipage}%
\begin{minipage}{.5\textwidth}
\centering
\includegraphics[width=.9\linewidth]{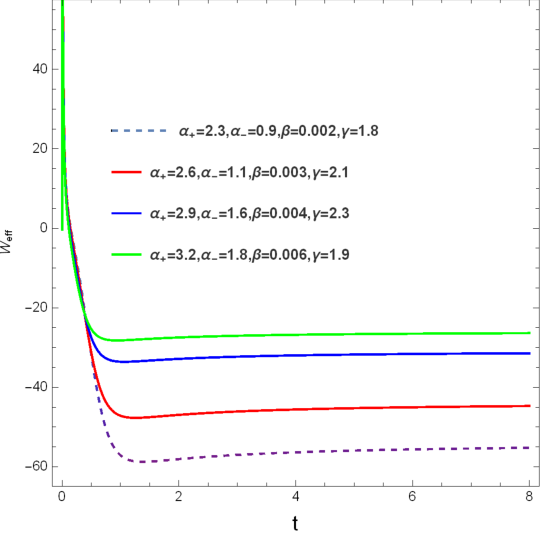}
\caption {Evolution of reconstructed viscous interacting  4-parameter entropic GHDE effective EoS parameter for the range $b~\text{lies in}~(0,1)$ in hybrid-scale factor scenario.}
\label{Fig_02_GHDE_weff}
\end{minipage}
\end{figure}
In the 4-parameter entropic most generalized viscous coupled GHDE scenario, Fig\ref{Fig_01_GHDE_w} shows the viscous interacting EoS parameter and Fig\ref{Fig_02_GHDE_weff} shows the effective EoS parameter of the reconstructed equivalent entropic dark energy scenario. In Fig\ref{Fig_01_GHDE_w}, the EoS parameter exhibits its phantom behavior and, in late time, EoS parameter is asymptotic to phantom boundary, while in Fig\ref{Fig_2_NO_weff}, the effective EoS parameter is rapidly approaching the phantom boundary, without overcoming the Big-rip singularity.
In this instance, we have investigated the stability of the model in an interacting most generalized coupled GHDE scenario, where $C_s^2=\frac{d(P_{effective})}{d(\rho_{effective})} \ge 0$ denotes the stability of the model under small perturbations.
\begin{figure}
\begin{minipage}{.5\textwidth}
  \centering
\includegraphics[width=.9\linewidth]{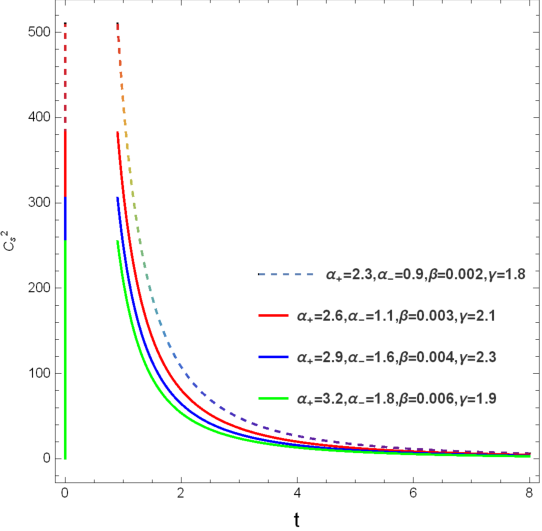}
\caption {Square speed of sound $C_s^2\ge 0$ in late-time for viscous coupled most generalized 4-parameter entropic GHDE in hybrid-scale factor scenario.}
\label{Fig_03_GHDE_Cs-square}
\end{minipage}
\end{figure}
It is found that our model is stable in the early and late times of the universe with background fluid as GHDE in a 4-parameter entropic interacting viscous hybrid-scale factor scenario utilizing the square speed of sound test in Fig\ref{Fig_03_GHDE_Cs-square}.

\subsection{Cosmological Settings of Generalized Holographic Dark fluid with Nojiri-Odintsov Cut-Off}
To get knowledge about the equation of state of holographic dark energy, we will look at the study \cite{HDE NO}. Furthermore, we will address a specific model of generalized holographic dark energy with the Nojiri-Odintsov cut-off combination, known as NOHDE. In order to develop NOHDE, we have thought of the following combination in Eq. (\ref{HDE}):
\begin{equation}
\frac{\zeta}{\mathcal{L}}=\frac{1}{R_h}[\tau_0 + \tau_1 R_h + \tau_2 R_h^2 ];
\label{infrared}
\end{equation}
where, $R_h=\frac{1}{H}$ is chosen as Hubble horizon and $\zeta$, $\tau_0$, $\tau_1$, and $\tau_2$ are the numerical constants. The feasibility of uniting the early- and late-time universes based on phantom cosmology is illustrated in \cite{HDE NO}, where a thorough examination of the rationale behind the consideration of such cut-offs is found. The notion of phantom non-phantom transition, which manifests in such a way that the universe could have effectively phantom equation of state at early-time as well as at late-time, is also one of the intriguing conclusions (among the others) described in \cite{HDE NO}. In general, there can be multiple phantom and non-phantom phases in the oscillating universe. To derive the Hubble horizon, we have chosen a scale factor in this regard. The logamediate scale factor, which has the form \cite{logamediate1, logamediate2}, is the subsequent one taken into consideration, 
\begin{equation}
a(t)=e^{\alpha (\log t)^{\beta }},
\label{lagamediate-scale factor}
\end{equation}
where, $\alpha>0$ and, $\beta>1$. Therefore, Hubble parameter (H) can take the form of
\begin{equation}
H=\frac{\alpha  \beta  (\log t)^{-1+\beta }}{t},
\label{Hubble lagamediate-scale factor}
\end{equation}
As we have chosen Hubble horizon in this case, 
\begin{equation}
R_h=\frac{1}{H},
\label{RH}
\end{equation}
and, $\mathcal{L}=\frac{\zeta R_H}{\tau _0+\tau _1R_H+\tau _2R_H^2}$ are our choices, we can solve for $\rho_{HDE}=\frac{3 \zeta ^2}{\mathcal{L}^2}$ in this reconstructed NOHDE scenario.
\begin{equation}
\rho_{NOHDE}=\frac{3 \alpha ^2 \beta ^2 (\log t)^{-2+2 \beta } \left(\tau _0+\frac{t (\log t)^{1-\beta } \tau _1}{\alpha  \beta }+\frac{t^2 (\log t)^{2-2 \beta } \tau _2}{\alpha ^2 \beta^2}\right)^2}{t^2}
\label{density NOHDE}
\end{equation}
Therefore, from Eqs. (\ref{Hubble lagamediate-scale factor}), (\ref{conservation1}) and $Q=3Hb^2\rho_{DM}$, we can get the expression of dark matter density in this reconstruction scheme;
\begin{equation}
\rho_{DM}=e^{3 \left(-1+b^2\right) \alpha  (\log t)^{\beta }} C_1.
\label{density-darkmatter-NOHDE}
\end{equation}
Hence, effective coupled viscous NOHDE dark fluid density will be;
\begin{equation}
\rho_{effective}=C_1 e^{3 \left(-1+b^2\right) \alpha  (\log t)^{\beta }}+\frac{3 (\log t)^{-2 (1+\beta )} \left(\alpha ^2 \beta^2 (\log t)^{2 \beta } \tau _0+t (\log t) \left(\alpha  \beta  (\log t)^{\beta } \tau _1+t (\log t) \tau _2\right)\right)^2}{t^2\alpha ^2 \beta ^2}
\label{total density-NOHDE}
\end{equation}
Therefore, from Eq. (\ref{2nd-Fried}) we can get the thermodynamic pressure of NOHDE and from Eq. (\ref{Hubble lagamediate-scale factor}) and $\Pi=-3 \xi H^2$ we can get the bulk viscous pressure in this NOHDE reconstructed scenario. Then, EoS parameter $w=\frac{P_{HDE}}{\rho_{effective}}$ and effective EoS parameter $w_{effective}=\frac{P_{HDE}+\Pi}{\rho_{effective}}$ can be expressed as:
\begin{equation}
\begin{array}{c}
w=
\frac{\mathcal{M}}{\left(3 \alpha  \beta  \left(\alpha ^2 \beta ^2 (\log t)^{2 \beta } \tau _0+t (\log t) \left(\alpha  \beta  (\log t)^{\beta }\tau _1+t (\log t) \tau _2\right)\right)^2\right)},
\end{array}
\label{w-NOHDE}
\end{equation}
where, 
\begin{equation*}
\begin{array}{c}
\mathcal{M}=((\log t)^{-\beta } (-\alpha^4 \beta^4 (\log t)^{4 \beta } \left(-2+2 \beta -2 (\log t)+3 \alpha  \beta  (\log t)^{\beta}\right) \tau _0^2-2 t \alpha ^3 \beta ^3 (\log t)^{1+3 \beta } \tau _0 \\((-1+\beta -(\log t)+3 \alpha  \beta  (\log t)^{\beta
}) \tau _1+3 t (\log t) \tau _2)-
t^2 (\log t)^2 (b^2 C_1 e^{3 \left(-1+b^2\right) \alpha  (\log t)^{\beta }} \alpha ^3 \beta ^3 (\log t)^{3\beta }\\+3 \alpha ^3 \beta ^3 (\log t)^{3 \beta } \tau _1^2+2 t \alpha  \beta  (\log t)^{1+\beta } \left(1-\beta +(\log t)+3 \alpha\beta  (\log t)^{\beta }\right) \tau _1 \tau _2\\+t^2 (\log t)^2 \left(2-2 \beta +2 (\log t)+3 \alpha  \beta  (\log t)^{\beta}\right) \tau_2^2))),
 \end{array}
\end{equation*}
and,
\begin{equation}
w_{effective}=-\frac{C_1 e^{3 \left(-1+b^2\right) \alpha  (\log t)^{\beta }}+\frac{6 \alpha  \beta  (\log t)^{-2+\beta } \left(-1+\beta-(\log t)+\alpha  \beta  (\log t)^{\beta }\right)}{t^2}+\mathcal{N}}{3 \left(C_1 e^{3 \left(-1+b^2\right) \alpha  (\log t)^{\beta }}+\frac{3 (\log t)^{-2 (1+\beta )} \left(\alpha ^2 \beta ^2 (\log t)^{2\beta } \tau _0+t (\log t) \left(\alpha  \beta  (\log t)^{\beta } \tau _1+t (\log t) \tau _2\right)\right)^2}{t^2 \alpha ^2 \beta^2}\right)}
\label{weffective-NOHDE}
\end{equation}
where,
\begin{equation*}
\mathcal{N}=\frac{3 (\log t)^{-2 (1+\beta )} \left(\alpha ^2 \beta ^2 (\log t)^{2\beta } \tau _0+t (\log t) \left(\alpha  \beta  (\log t)^{\beta } \tau _1+t (\log t) \tau _2\right)\right)^2}{t^2 \alpha ^2 \beta^2}.
\end{equation*}
\begin{figure}
\centering
\begin{minipage}{.5\textwidth}
  \centering
\includegraphics[width=.9\linewidth]{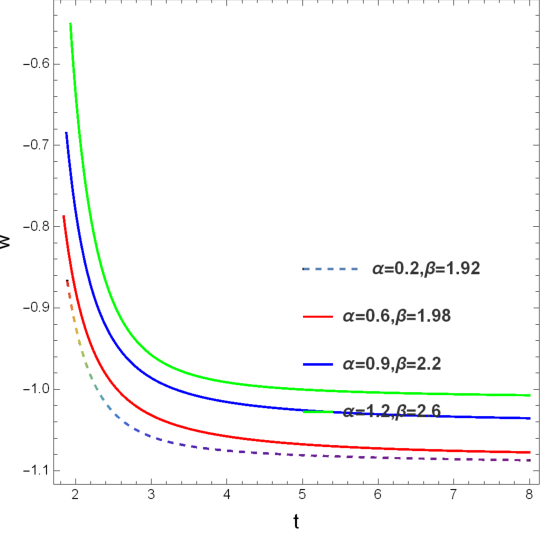}
\caption{Evolution of reconstructed viscous interacting NOHDE EoS parameter for the range $b~\text{lies in}~(0,1)$ in logamediate scenario.}
\label{Fig_1_NO_w}
\end{minipage}%
\begin{minipage}{.5\textwidth}
\centering
\includegraphics[width=.9\linewidth]{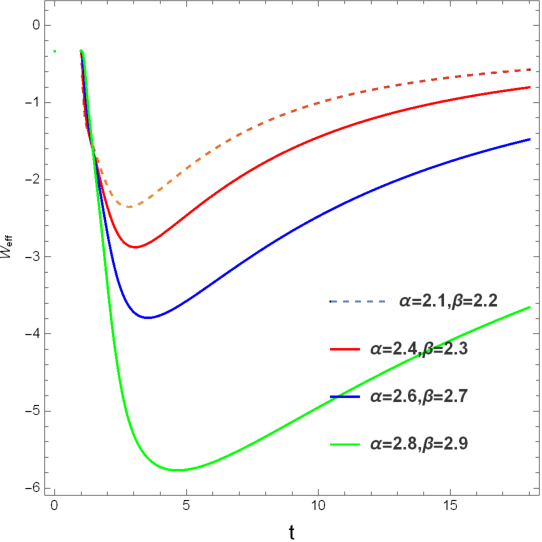}
\caption {Evolution of reconstructed viscous interacting NOHDE effective EoS parameter for the range $b~\text{lies in}~(0,1)$ in logamediate scenario.}
\label{Fig_2_NO_weff}
\end{minipage}
\end{figure}
In the logamediate case, Fig\ref{Fig_1_NO_w} shows the viscous interacting EoS parameter and Fig\ref{Fig_2_NO_weff} shows the effective EoS parameter of the reconstructed NOHDE. In Fig\ref{Fig_1_NO_w}, the EoS parameter exhibits its quintessential behavior and, in late time, is crossing the phantom boundary, while in Fig\ref{Fig_2_NO_weff}, the effective EoS parameter is steadily approaching the phantom boundary, although it can overcome the Big-rip singularity.
Here, we have examined the model's stability in an interaction situation, where $C_s^2 \ge 0$ indicates that the model is stable under small perturbations.
\begin{figure}
\begin{minipage}{.5\textwidth}
  \centering
\includegraphics[width=.9\linewidth]{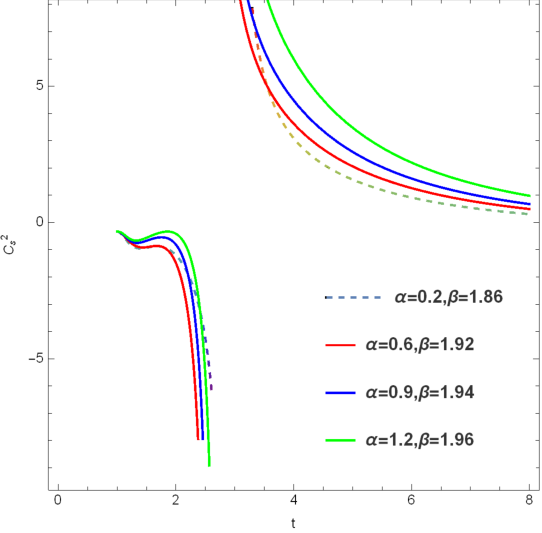}
  \caption {Square speed of sound $C_s^2\ge 0$ in late-time for viscous NOHDE in interacting logamediate scenario.}
\label{Fig_3_Cs-square}
\end{minipage}
\end{figure}
By using the square speed of sound test in Fig\ref{Fig_3_Cs-square}, it is determined that our model is unstable in the early stages of the universe but stable against small perturbations in the late-time with background fluid as NOHDE in an interacting viscous logamediate scenario.
\subsection{Cosmological Settings of Tsallis Entropic Dark Fluid}
\begin{figure}
\begin{minipage}{.5\textwidth}
\centering
\includegraphics[width=.9\linewidth]{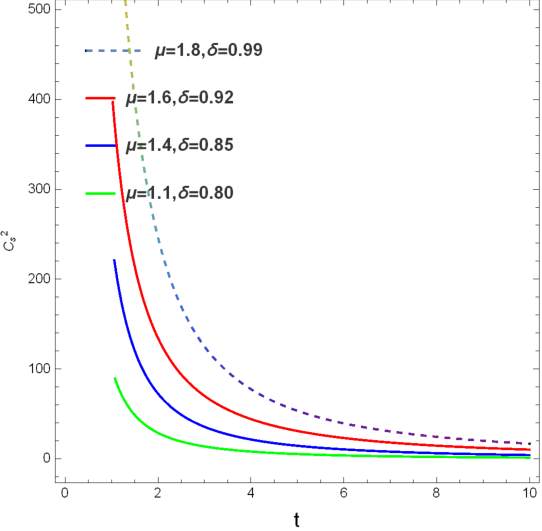}
\caption {Square speed of sound $C_s^2\ge 0$ in late-time for viscous THDE in interacting intermediate scenario.}
\label{Fig_6_Tsallis_Cs-square}
\end{minipage}
\end{figure}

The THDE model is based on the modified entropy-area relation \cite{HDE TSallis1}, and the following energy density was introduced in \cite{HDE TSallis2} to suggest the holographic dark energy hypothesis;
\begin{equation}
\rho_{THDE} = B \mathcal{L}^{(2\delta-4)},
\label{Tsallis density}
\end{equation}
where $\delta$ is the non-additivity parameter and $B$ is an unknown parameter \cite{HDE TSallis1,HDE TSallis2}. It is important to note that the foregoing connection yields the typical HDE $\rho_{DE} = B \mathcal{L}^{-2}$ in the particular situation $\delta=1$, where $B = 3c^2 M_p^2$, and $c^2$ and $M_p^2$ stand for the model parameter and reduced Planck mass, respectively. Furthermore, Eq. (\ref{Tsallis density}) provides the usual cosmological constant case $\rho_{DE} =\Lambda= Constant$ for $\delta = 2$. As a result, THDE is a more general framework than the conventional HDE. From this point on, we only address the general case, which includes $\delta = 1$ and $\delta = 2$. Since the positive body of DE itself is still unknown at this point, we suggest applying holography and entropy relations to the entire universe—a gravitational, non-extensive system—as a potential method of obtaining the hint that would reveal the nature of DE. Specifically, we employ the generalized definition of the universe horizon entropy found in  Eq. (\ref{Tsallis density}) and discuss the Tsallis entropy in this work. The future event horizon is the most commonly used one in the literature, although there are other possibilities for the greatest length $\mathcal{L}$ that occurs in the expression of any holographic dark energy. 
\begin{equation}
R_E=a \int_{t}^{\infty}\frac{dt}{a}=a \int_{a}^{\infty}\frac{da}{H a^2}
\label{future-event}
\end{equation}
where, $a$ denotes scale factor and $H$ be the Hubble parameter. Hence, by replacing $\mathcal{L}$ in equation Eq. (\ref{Tsallis density}) with $R_E$, we get to the energy density of THDE, which is,
\begin{equation}
\rho_{THDE} = B R_E^{(2\delta-4)},
\label{Tsallis density2}
\end{equation}
In this regard, we have chosen scale factor as an intermediate scale factor\cite{intermediate1,intermediate2} with 
\begin{equation}
a=e^{\mu  t^n},
\label{intermediate scalefactor}
\end{equation}
where, $\mu>0$ and $n\in (0,1)$. Therefore, Hubble parameter can be rewritten as;
\begin{equation}
H=n t^{-1+n} \mu 
\label{Hubble intermediate}
\end{equation}
Hence, future event horizon can be written as in this reconstructed scenario:
\begin{equation}
R_E=e^{t^n \mu } \left(C_1+\frac{t \left(t^n \mu \right)^{-1/n} \Gamma\left[\frac{1}{n},t^n \mu \right]}{n}\right).
\label{RE}
\end{equation}
Therefore, Eqs. (\ref{Tsallis density2}) and (\ref{RE}), we can easily find the Tsallis energy density in this case;
\begin{equation}
\rho_{THDE}=B \left(e^{t^n \mu } \left(C_1+\frac{t \left(t^n \mu \right)^{-1/n} \Gamma\left[\frac{1}{n},t^n \mu \right]}{n}\right)\right)^{2\delta -4}.
\label{Tsallis darkenergy density}
\end{equation}
Again, from Eqs. (\ref{conservation1}) and (\ref{Hubble intermediate}), we can get the interacting dark-matter and effective energy density respectively in this case;
\begin{equation}
\rho_{DM}=e^{3 \left(-1+b^2\right) t^n \mu } C_2,
\label{Tsallis darkmatter density}
\end{equation}
\begin{equation}
\rho_{effective}=C_2 e^{3 \left(-1+b^2\right) t^n \mu }+B \left(e^{t^n \mu } \left(C_1+\frac{t \left(t^n \mu \right)^{-1/n} \Gamma\left[\frac{1}{n},t^n
\mu \right]}{n}\right)\right)^{2 (-2+\delta )}.
\label{total density TSallis}
\end{equation}
Thus, in this THDE reconstructed scenario, we can obtain the thermodynamic pressure (P) of THDE from Eq. (\ref{2nd-Fried}) by putting effective THDE density, and the bulk viscous pressure from Eq. (\ref{Hubble intermediate}) and $\Pi=-3 \xi H^2$. Therefore, reconstructed EoS parameter and effective EoS parameter can be written as;
\begin{equation}
\begin{array}{c}
w=\frac{1}{3 B n \mu }t^{1-n} \left(e^{t^n \mu } \left(C_1+\frac{t \left(t^n \mu \right)^{-1/n} \Gamma\left[\frac{1}{n},t^n \mu
\right]}{n}\right)\right)^{4-2 \delta }
(-3 b^2 C_2 e^{3 \left(-1+b^2\right) t^n \mu } n t^{-1+n} \mu\\ -3 B n t^{-1+n} \mu  \left(e^{t^n \mu } \left(C_1+\frac{t \left(t^n
\mu \right)^{-1/n} \Gamma\left[\frac{1}{n},t^n \mu \right]}{n}\right)\right)^{-4+2 \delta }-B (-4+2 \delta ) \left(e^{t^n \mu } \left(C_1+\frac{t \left(t^n \mu \right)^{-1/n} \Gamma\left[\frac{1}{n},t^n \mu \right]}{n}\right)\right)^{-5+2
\delta }\\ \left(e^{t^n \mu } \left(-e^{-t^n \mu }-\frac{t^n \mu  \left(t^n \mu \right)^{-1-\frac{1}{n}} \Gamma\left[\frac{1}{n},t^n \mu \right]}{n}+\frac{\left(t^n
\mu \right)^{-1/n} \Gamma\left[\frac{1}{n},t^n \mu \right]}{n}\right)+e^{t^n \mu } n t^{-1+n} \mu  \left(C_1+\frac{t \left(t^n \mu \right)^{-1/n}
\Gamma\left[\frac{1}{n},t^n \mu \right]}{n}\right)\right)),
\end{array}
\label{w-Tsallis}
\end{equation}
\begin{equation}
\begin{array}{c}
w_{eff}=-\frac{\mathcal{P}}{C_2 e^{3 \left(-1+b^2\right) t^n \mu }+B \left(e^{t^n \mu } \left(C_1+\frac{t
\left(t^n \mu \right)^{-1/n} \Gamma\left[\frac{1}{n},t^n \mu \right]}{n}\right)\right)^{2 (-2+\delta )}},
\end{array}
\label{w-Tsallis}
\end{equation}
where,
\begin{equation}
\begin{array}{c}
\mathcal{P}=3 n^2 t^{-2+2 n} \mu ^2 \xi +\frac{1}{3} (C_2 e^{3 \left(-1+b^2\right) t^n \mu }+6 n t^{-2+n} \mu  \left(-1+n+n
t^n \mu \right)\\-9 n^2 t^{-2+2 n} \mu ^2 \xi +B \left(e^{t^n \mu } \left(C_1+\frac{t \left(t^n \mu \right)^{-1/n} \Gamma\left[\frac{1}{n},t^n
\mu \right]}{n}\right)\right)^{2 (-2+\delta )}).
\end{array}
\end{equation}
\begin{figure}
\centering
\begin{minipage}{.5\textwidth}
  \centering
  \includegraphics[width=.9\linewidth]{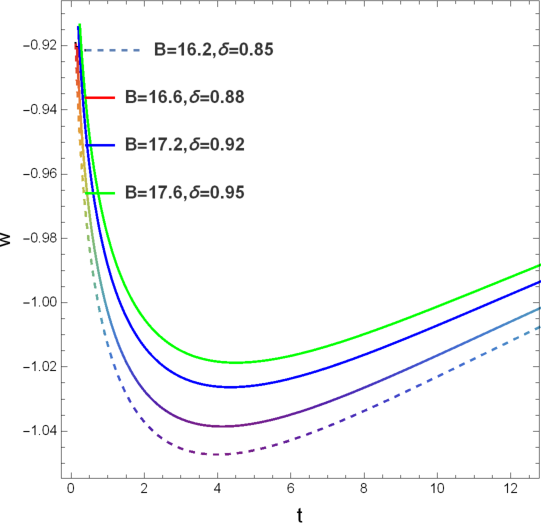}
  \caption{Evolution of reconstructed viscous interacting THDE EoS parameter with the range of interaction coefficient $b~\text{lies in}~(0,1)$ in intermediate scenario.}
  \label{Fig_4_Tsallis_w}
\end{minipage}%
\begin{minipage}{.5\textwidth}
  \centering
  \includegraphics[width=.9\linewidth]{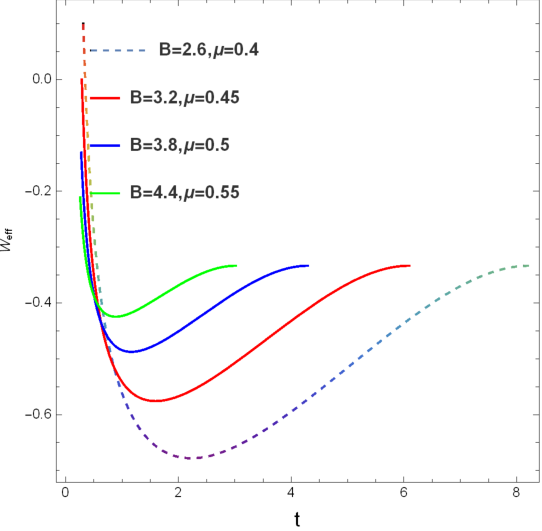}
  \caption {Evolution of reconstructed viscous interacting effective THDE EoS parameter with the range of interaction coefficient $b~\text{lies in}~(0,1)$ in intermediate scenario.}
  \label{Fig_5_Tsallis_weff}
\end{minipage}
\end{figure}
In the interacting viscous scenario, we obtained the expression of the EoS parameter $w$, which we showed against time in Fig{\ref{Fig_4_Tsallis_w}}, where, we obtained the expression of pressure $(P)$ by placing the expressions of $\rho_{effective}$ and $\Pi$ in Eq. (\ref{2nd-Fried}). In Fig{\ref{Fig_5_Tsallis_weff}}, we have shown $w_{eff}=\frac{P+\pi}{\rho_{Total}}$ in an interacting viscous reconstructed THDE scenario. Now, we have examined the model's stability in this interacting situation, where $c_s^2 \ge 0$ indicates that the model is stable under mild perturbations. Early in the time series, it is observed that the interacting EoS parameter $w$ in Fig{\ref{Fig_4_Tsallis_w}} crosses the $-1$ phantom barrier. After that, it can avoid big-rip singularity and exhibit quintessential behavior in the later phase time. The interacting effective EoS parameter in Fig{\ref{Fig_5_Tsallis_weff}} acts as a quintessence.  Fig{\ref{Fig_6_Tsallis_Cs-square}} shows our model in visocus interacting THDE scenario is stable under small perturbations.
\subsection{Cosmological Settings of Barrow Entropic Dark Fluid}
\begin{figure}
\centering
\begin{minipage}{.5\textwidth}
  \centering
  \includegraphics[width=.9\linewidth]{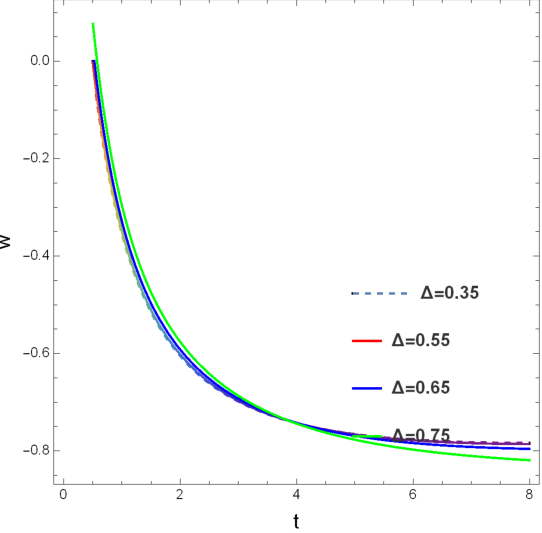}
  \caption{Evolution of reconstructed viscous interacting BHDE EoS parameter with the range of quantum gravitational deformation parameter $\Delta~\text{lies in}~(0,1)$ in emergent scenario.}
  \label{Fig_7_BHDE_w}
\end{minipage}%
\begin{minipage}{.5\textwidth}
  \centering
  \includegraphics[width=.9\linewidth]{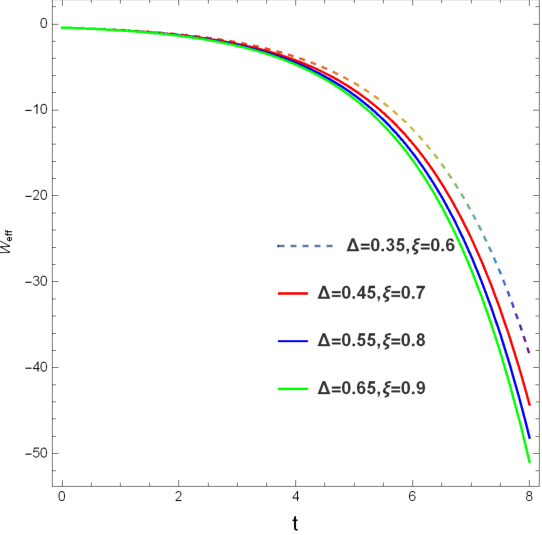}
  \caption {Evolution of reconstructed viscous interacting effective BHDE EoS parameter with the range of quantum gravitational deformation parameter $\Delta~\text{lies in}~(0,1)$ in emergent scenario.}
  \label{Fig_8_BHDE_weff}
\end{minipage}
\end{figure}

\begin{figure}
\begin{minipage}{.5\textwidth}
\centering
\includegraphics[width=.9\linewidth]{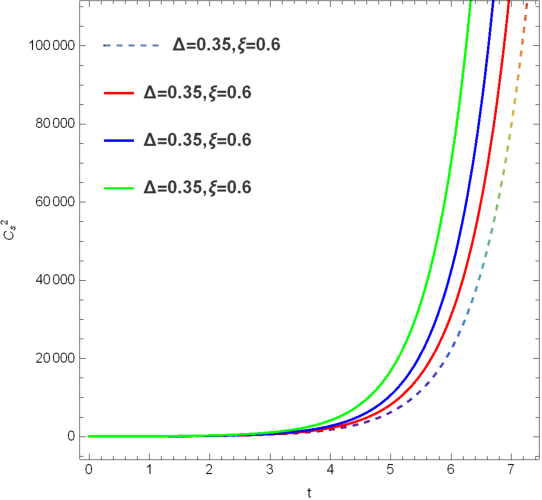}
\caption {Square speed of sound $C_s^2\ge 0$ in late-time for viscous BHDE in interacting emergent scenario.}
\label{Fig_9_BHDE_Cs-square}
\end{minipage}
\end{figure}

This section examines Barrow holographic dark energy (BHDE) in the context of bulk viscosity. Taking into account the entropy condition $\mathcal{S} \propto A \propto \mathcal{L}^2$ with the standard entropy relation $\rho_{DE}\mathcal{L}^4\le \mathcal{S}$, where $\mathcal{L}$ denotes the horizon's length and $\rho_{DE}$ represents the standard holographic dark energy. The energy density of Barrow Holographic Dark Energy (BHDE) can be obtained using the black-hole entropy formula  $\mathcal{S}_h=\left(\frac{A}{A_0}\right)^{1+\frac{\Delta}{2}}$. It looks like this: 
\begin{equation} 
\rho_{BHDE}=c \mathcal{L}^{\Delta-2} 
\label{density1 BHDE} 
\end{equation}
In this case, $c=3p^2M_p^2$, where $p^2$ is our model parameter, $\Delta$ represents quantum gravitational deformation, and $\mathcal{L}$ is the length of the infrared cut-off. $M_p$ stands for reduced Planck mass. The conventional holographic dark energy which is expected when $\Delta=0$ is provided by the above suggested  expression of BHDE. Nevertheless, by changing the range of $\Delta$, Barrow holographic dark energy will deviate from the traditional one and exhibit unique cosmic behavior. The maximum length $\mathcal{L}$ that occurs in the expression of any holographic dark energy can have other values, but the future event horizon is the one that is most commonly employed in the literature \cite{Barrow2}. In this case, we have considered particle horizon as length of the IR cut-off which is;
\begin{equation}
R_P= a(t) \int_{0}^{t}\frac{dt}{a(t)},
\label{particle horizon}
\end{equation}
and, \begin{equation}
\dot{R_P}=HR_P+1
\end{equation}
For calculation particle horizon at this moment we have choose scale factor in this reconstructed scenario. Here, we have chosen emergent scale factor\cite{emergent1,emergent2} for this reconstruction;
\begin{equation}
a=a_0 \left(\lambda +e^{\mu  t}\right)^n,
\label{emergent scale factor}
\end{equation}
wherein $n$, $\mu$, $\lambda$, and $a_0$ are positive invariants. The scale factor $a(t)$ must be positive, $a_0>0$ must be satisfied, and $\lambda$ must be positive in order to prevent big-rip singularities. Both $n$ and the scale factor $a(t)$ need to be positive for the universe to expand more quickly. $t=-\infty$ is the big-bang singularity that we must reach if $a<0$ and $n<0$. Thus, using Eq. (\ref{emergent scale factor}), we can determine the Hubble parameter (H), which is as follows:
\begin{equation}
H=\frac{e^{t \mu } n \mu }{e^{t \mu }+\lambda}
\label{Hubble-emergent}
\end{equation}
Hence,
\begin{equation}
R_P=\left(e^{t \mu }+\lambda \right)^n C_1-\frac{\left(1+e^{-t \mu } \lambda \right)^n  {}_2F_1\left[n,n,1+n,-e^{-t \mu} \lambda \right]}{n \mu }
\label{RP}
\end{equation}
Therefore, from Eqs. (\ref{Hubble-emergent}) and (\ref{RP}), BHDE energy density $\rho_{BHDE}=c R_P^{\Delta -2}$, can be rewritten as:
\begin{equation}
\rho_{BHDE}=c \left(\left(e^{t \mu }+\lambda \right)^n C_1-\frac{\left(1+e^{-t \mu } \lambda \right)^n  {}_2F_1\left[n,n,1+n,-e^{-t\mu } \lambda \right]}{n \mu }\right)^{-2+\Delta }
\label{BHDE darkenergy density}
\end{equation}
From Eq. (\ref{conservation2}), we got the solutions of matter density in this viscous interacting BHDE scenario;
\begin{equation}
\rho_{DM}=\left(e^{t \mu }+\lambda \right)^{3 \left(-1+b^2\right) n} C_2.
\label{BHDE darkmatter density}
\end{equation}
Therefore,
\begin{equation}
\rho_{effective}=C_2 \left(e^{t \mu }+\lambda \right)^{3 \left(-1+b^2\right) n}+c \left(C_1 \left(e^{t \mu }+\lambda \right)^n-\frac{\left(1+e^{-t\mu } \lambda \right)^n  {}_2F_1\left[n,n,1+n,-e^{-t \mu } \lambda \right]}{n \mu }\right)^{-2+\Delta}
\label{BHDE total density}
\end{equation}
The thermodynamic pressure (P) expression can be obtained from Eq. (\ref{conservation2}), and the EoS parameter $w$ expression can be obtained from Eqs. (\ref{conservation2}), and (\ref{BHDE darkmatter density}) and plotted it in Fig\ref{Fig_7_BHDE_w}. It is known that the effective EoS parameter $w_{eff}$, 
$w_{eff}=\frac{P+\Pi}{\rho_{Total}}$, we can calculate,
\begin{equation}
\begin{array}{c}
w_{eff}=-\frac{\mathcal{X}}{3 \left(e^{t \mu }+\lambda
\right)^2 \left(C_2 \left(e^{t \mu }+\lambda \right)^{3 \left(-1+b^2\right) n}+c \left(C_1 \left(e^{t \mu }+\lambda \right)^n-\frac{\left(1+e^{-t
\mu } \lambda \right)^n  {}_2F_1\left[n,n,1+n,-e^{-t \mu } \lambda \right]}{n \mu }\right)^{-2+\Delta }\right)}
\end{array}
\end{equation}
where,
\begin{equation}
\begin{array}{c}
\mathcal{X}=C_2 \left(e^{t \mu }+\lambda \right)^{2+3 \left(-1+b^2\right) n}-6 e^{2 t \mu } n \mu ^2+6 e^{2 t \mu } n^2 \mu ^2+6
e^{t \mu } n \left(e^{t \mu }+\lambda \right) \mu ^2 \\ +c \left(e^{t \mu }+\lambda \right)^2 \left(C_1 \left(e^{t \mu }+\lambda \right)^n-\frac{\left(1+e^{-t
\mu } \lambda \right)^n  {}_2F_1\left[n,n,1+n,-e^{-t \mu } \lambda \right]}{n \mu }\right)^{-2+\Delta }
\end{array}
\end{equation}
The expression of the EoS parameter $w$ in the interacting viscous BHDE scenario is derived and is plotted against time in Fig{\ref{Fig_7_BHDE_w}} and effective EoS parameter in Fig{\ref{Fig_8_BHDE_weff}}. After looking at the model's stability in this viscous interacting BHDE scenario, we can see that it is stable under modest perturbations since $c_s^2 \ge 0$. The interacting EoS parameter $w$ in Fig{\ref{Fig_7_BHDE_w}} is seen to behave quintessential behavior. In Fig{\ref{Fig_8_BHDE_weff}}, the interacting effective EoS parameter serves as phantom behaviour without prevention of Big-rip singularity in later phases of universe. Our model in the viscous interacting BHDE scenario is stable under minor perturbations, as demonstrated by Fig{\ref{Fig_9_BHDE_Cs-square}}. 
\section{Generalized Second Law of Thermodynamics with Bekenstein Entropy of Interacting Different Faces of HDE}
In this section, we will examine the validity of generalized second law of thermodynamics (GSLT) after determining the rate of change of total entropy. The principles of thermodynamics that apply to black holes also apply to cosmic horizons, as is well known. Since it is commonly acknowledged that a cosmic horizon shares many of the same thermodynamic properties as a black hole, studying the thermodynamics of the gravity theory is an interesting field of study. Moreover, the first Friedmann equation in the FRW universe can also be used to find the first law of thermodynamics, which holds at a black hole horizon, when the universe is constrained by an apparent horizon. Bekenstein\cite{thermodynamic2} proposed in 1973 that the thermodynamics and event of horizon of a black hole are related, i.e., the event of horizon of a black hole is a measure of its entropy. We have extended this idea to cosmological model horizons, where each horizon corresponds to an entropy. The second law of thermodynamics was thus changed in a way that required the total of entropies to rise with time, as stated in its extended form: the sum of all horizon-related time derivatives of entropies plus the time derivative of normal entropy must be positive. This provides a compelling argument for utilizing the future event horizon as the cosmic horizon when examining the thermodynamic properties of any cosmological model. Here, we have considered the universe to be a thermodynamic system, bounded by the cosmic event horizon of radius\cite{thermodynamic36,thermodynamic37}, in light of the previously cited arguments. Considering the previously cited points of view, we have considered the universe to be a thermodynamic system in this instance, bounded by the cosmic aforementioned horizons with the respective background fluids. Following the clarification of the black hole's thermodynamic properties \cite{thermodynamic1,thermodynamic2,thermodynamic3}, it has been stated that the black hole's entropy is proportional to the area $A$ of the horizon, which is:
\begin{equation}
\mathcal{S}_h=\frac{A}{4G},~\text{where},~A=4\pi R_h^2.
\label{Bekenstein-area}
\end{equation}
Let us consider the overall entropy $(\mathcal{S})$, which will be $\mathcal{S}=\mathcal{S}_h+\mathcal{S}_f$, where the entropies of the fluid and the horizon it includes are denoted, respectively, by $\mathcal{S}_f$ and $\mathcal{S}_h$. For every solitary macroscopic system, $\mathcal{S}$ must satisfy the following relations, which are consistent with the principles of thermodynamics. Thus, by virtue of thermodynamic equilibrium (TE) and the second law of thermodynamics (GSL),
 \begin{equation}
\begin{array}{c}
\dot{\mathcal{S}}=\frac{d\mathcal{S}}{dt}\ge0~~and~~\ddot{\mathcal{S}}=\frac{d^2\mathcal{S}}{dt^2}<0.
\end{array}
\label{sdot}
\end{equation}
We have taken into account both viscous interacting dark matter and various HDE forms in order to verify the validity of GSL. Assuming that the horizons and the dark sectors are in equilibrium—that is, that the horizon's temperature is the same for dark energy and dark matter—we evaluate the viability of GSL. The temperature near each of horizon is
\begin{equation}
\begin{array}{c}
\mathcal{T}_E=\frac{1}{2 \pi R_h}
\end{array}
\label{horizon temperature}
\end{equation}
As previously mentioned, we considered the viscous couple different faces of HDE and DM as the components of the background fluid, then we can write,
\begin{equation}
\mathcal{S}_f=\mathcal{S}_{HDE}+\mathcal{S}_{DM}
\label{entropy-fluid}
\end{equation}
where the temperature of this single fluid inside the horizon is denoted by $\mathcal{T}$, and the entropies of the interacting fluids of HDE forms and DM are indicated by $\mathcal{S}_{HDE}$ and $\mathcal{S}_{DM}$, respectively. Consequently, the first law of thermodynamics $(\mathcal{T}d\mathcal{S} = dE + PdV)$ can be written as follows for the individual matter contents:
\begin{equation}
\mathcal{T}d\mathcal{S}_{HDE} = dE_{HDE} + P_{HDE} dV,
\label{TdS_{HDE}}
\end{equation}
\begin{equation}
\mathcal{T}d\mathcal{S}_{DM} = dE_{DM} + P_{DM}dV = dE_{DM},~\text{as} ~~P_{DM}=0
\label{TdS_{DM}}
\end{equation}
where the expression $V = \frac{4}{3} \pi R_h^3$ represents the horizon volume. The internal energies of pressure-less DM and HDE forms are again represented by the expressions $E_{DM}=\frac{4}{3} \pi R_h^3\rho_{DM}$ and $E_{HDE}=\frac{4}{3} \pi R_h^3 \rho_{HDE}$. Consequently, we can obtain $\dot{\mathcal{S}}_{HDE}$ and $\dot{\mathcal{S}}_{DM}$ by differentiating Eqs. (\ref{TdS_{HDE}}) and (\ref{TdS_{DM}}). It's also critical to keep in mind that, in this particular case, if the fluid temperature $\mathcal{T}$ deviates from the horizon temperature $\mathcal{T}_h$, the geometry would deform as a result of energy flow. As a result, inside this non-interacting HDE and DM, the entropy will be:
\begin{equation}
\begin{array}{c}
\dot{\mathcal{S}}_f=\dot{\mathcal{S}}_{HDE}+\dot{\mathcal{S}}_{DM}
\end{array}
\label{entropy-fluid}
\end{equation}
Again, entropy of the horizon can be defined as:
\begin{equation}
\begin{array}{c}
\mathcal{S}_h=\frac{A}{4G}, ~\text{where,}~A=4\pi R_h^3
\end{array}
\label{entropy-horizon}
\end{equation}
Here, $A=4\pi R_h^3$ be the surface area and $R_h$ be the radius of the respective horizons which we got from Eqs.(\ref{RH}), (\ref{RE}) and, (\ref{RP}) for the corresponding forms of coupled HDE as background fluid. Therefore, by differentiating Eq. (\ref{entropy-fluid}), we can arrive at the expression:
\begin{equation}
\begin{array}{c}
\dot{\mathcal{S}}_f= \dot{\mathcal{S}}_{HDE}+\dot{\mathcal{S}}_{DM}
\end{array}
\label{dot-entropy-fluid}
\end{equation}
Therefore, by 
\begin{equation}
\dot{\mathcal{S}}=\dot{\mathcal{S}}_h+\dot{\mathcal{S}}_{HDE}+\dot{\mathcal{S}}_{DM}
\label{time derivative of total entropy}
\end{equation}
\subsection{Generalized Second Law of Thermodynamics at Hubble Horizon}
\begin{figure}
\centering
\begin{minipage}{.5\textwidth}
  \centering
\includegraphics[width=.9\linewidth]{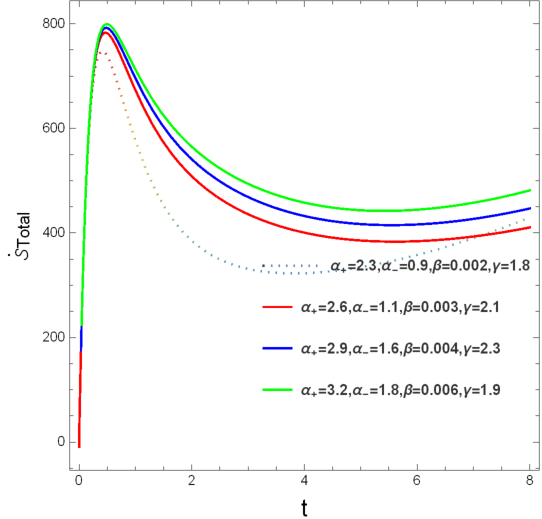}
\caption{Plot of $\dot{\mathcal{S}}$ to $t$ with time expressed in seconds, using Bekenstein entropy as entropy at the apparent horizon with the background fluid as reconstructed viscous coupled 4-parameter entropic most generalized holographic dark fluid.}
\label{Fig_04_GHDE_Sdot}
\end{minipage}%
\centering
\begin{minipage}{.5\textwidth}
  \centering
\includegraphics[width=.9\linewidth]{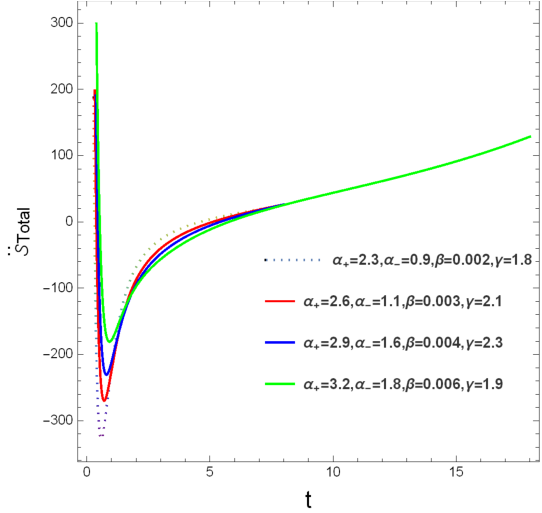}
\caption{Plot of thermal equilibrium $\ddot{\mathcal{S}}$ to $t$ with time expressed in seconds, using Bekenstein entropy as entropy at the apparent horizon  with the background fluid as reconstructed viscous coupled 4-parameter entropic most generalized holographic dark fluid.}
\label{Fig_05_GHDE_Sddot}
\end{minipage}%
\end{figure}

In this paper, we propose a study based on the 4-parameter most generalized entropy and the holographic dark energy reconstruction of the entire universe, a gravitational and non-extensive system, as a potential method to reveal the nature of DE. In particular, we employed the generalized definition of the universe horizon entropy found in Eq. (\ref{4-parameter entropy}) and discussed the 4-parameter most generalized entropy as well as the Nojiri-Odintsov cut-off, Tsalis entropy, and Barrow entropy in this work. It is expected that the Lagrangian for HDE can be written clearly and that we have found the equations of motion for such a physical system, which leads to the energy density of the DE component in Eq. (\ref{HDE}), if further research finds out how holography works physically and what the basic relationship is between gravity and thermodynamics. Despite the Lagrangian form, we were able to obtain the corresponding representation of the various types of HDE density using a phenomenological technique. This serves as both a good reason and a justification for looking at HDE models. For the 4-parameter generalized viscous coupled holographic reconstruction case, the Hubble horizon is taken into account in Eq. (\ref{infrared-four parameter}), that is, $R_H = H^{-1}$, the energy density and reflective cosmological parameters associated with GHDE are calculated, and now we are focusing on validation of GSLT in this regard. The second law of thermodynamics in the context of a cosmological system is defined as the sum of the entropies of all the constituents (mostly dark matter and DE) plus the entropy of the boundary of the universe (here, the Hubble horizon), which is always increasing. Thus, we can obtain the time derivative of total entropy in a viscous interacting 4-parameter entropic GHDE scenario by substituting $R_H$ from Eq. (\ref{R-H}), $\rho_{m}$ from Eq. (\ref{4-parameter dark-matter density}), $\rho_{Total}$ from Eq. (\ref{total density 4-parameter}), and thermodynamic pressure $P$ from Eq. (\ref{pressure 4-parameter}) and viscous pressure ($\Pi$) infused by coupled GHDE in Eqs. (\ref{entropy-fluid}), (\ref{entropy-horizon}). It is plotted in Fig\ref{Fig_04_GHDE_Sdot} Since the time derivative of total entropy fulfills the inequality $\dot{\mathcal{S}}\ge 0$, it is compatible with Fig\ref{Fig_04_GHDE_Sdot}. In addition, the thermal equilibrium (TE) for $\ddot{\mathcal{S}} \leq 0$ is established. 
As a result, Fig\ref{Fig_04_GHDE_Sdot} shows that $\dot{\mathcal{S}}$ remains positive as $t$ grows, supporting the validity of GSLT at the apparent horizon with Bekenstein entropy. By differentiating Eq. (\ref{time derivative of total entropy}), we extract provided $\ddot{\mathcal{S}}$ in order to examine the thermodynamic equilibrium. We plot $\ddot{\mathcal{S}}$ vs $t$ in Fig\ref{Fig_05_GHDE_Sddot} to show that the thermal equilibrium condition is satisfied for late-time pf the universe for Bekenstein entropy at apparent horizon.

\begin{figure}
\centering
\begin{minipage}{.5\textwidth}
  \centering
\includegraphics[width=.9\linewidth]{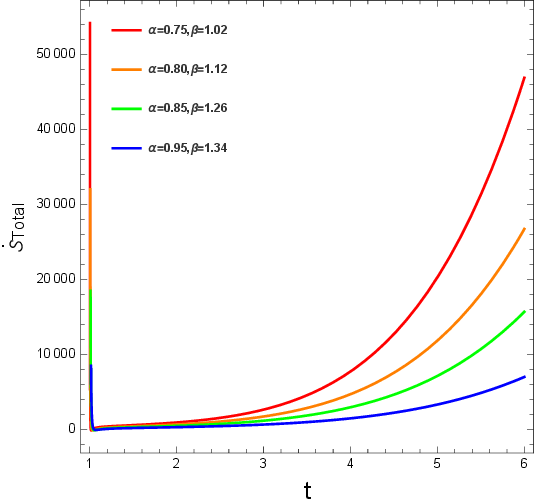}
\caption{Plot of $\dot{\mathcal{S}}$ to $t$ with time expressed in seconds, using Bekenstein entropy as entropy at the apparent horizon  with the background fluid as reconstructed viscous coupled generalized holographic dark fluid with Nojiri-Odintsov cut-off.}
\label{Fig_10_NOHDE_Sdot}
\end{minipage}%
\centering
\begin{minipage}{.5\textwidth}
  \centering
\includegraphics[width=.9\linewidth]{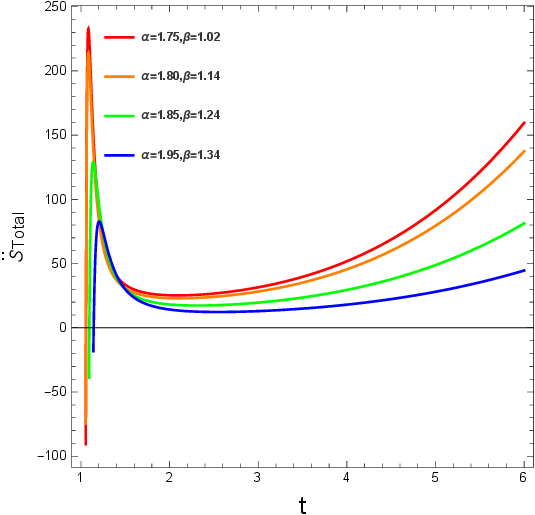}
\caption{Plot of thermal equilibrium $\ddot{\mathcal{S}}$ to $t$ with time expressed in seconds, using Bekenstein entropy as entropy at the apparent horizon  with the background fluid as reconstructed viscous coupled generalized holographic dark fluid with Nojiri-Odintsov cut-off.}
\label{Fig_11_NOHDE_Sddot}
\end{minipage}%
\end{figure}

We can estimate the amount of information behind certain horizons in cosmological models of an accelerated Universe by assigning them an entropy. The apparent horizon, also known as the Hubble horizon, has a radius of $R_A = R_H = \frac{1}{H}$, making it the most naturally occurring horizon in the universe.  We now verify the validity of thermodynamic equilibrium and GSLT for a physical system that is isolated and has a maximum entropy state. In the context of a cosmological system, the second law of thermodynamics is defined as the total of all entropies of the constituents (primarily dark matter and DE) and the entropy of the universe's boundary (in this case Hubble horizon), which can never decrease. Hence, by replacing $R_H$ from Eq. (\ref{RH}), $\rho_{m}$ from Eq. (\ref{density-darkmatter-NOHDE}), $\rho_{Total}$ from Eq. (\ref{total density-NOHDE}) and thermodynamic pressure $P$ and viscous pressure ($\Pi$) infused by coupled NOHDE in Eqs.(\ref{entropy-fluid}), (\ref{entropy-horizon}), we can get the time derivative of total entropy in viscous interacting NOHDE scenario. Plotting of it is shown in Fig\ref{Fig_10_NOHDE_Sdot}. The time derivative of total entropy is consistent with Fig\ref{Fig_10_NOHDE_Sdot} because, it satisfies the inequality $\dot{\mathcal{S}}\ge 0$. The thermal equilibrium (TE) is also established for $\ddot{\mathcal{S}} \leq 0$.
Hence, Fig\ref{Fig_10_NOHDE_Sdot} demonstrates that as $t$ increases, $\dot{\mathcal{S}}$ stays positive, proving the validity of GSLT at the apparent horizon with Bekenstein entropy. In order to investigate the thermodynamic equilibrium, we extract given $\ddot{\mathcal{S}}$ by differentiating Eq. (\ref{time derivative of total entropy}). In Fig\ref{Fig_11_NOHDE_Sddot}, we plot $\ddot{\mathcal{S}}$ vs $t$, demonstrating that for Bekenstein entropy at apparent horizon, the thermal equilibrium requirement is satisfied.

\subsection{Generalized Second Law of Thermodynamics at Future Event Horizon}
\begin{figure}
\centering
\begin{minipage}{.5\textwidth}
  \centering
\includegraphics[width=.9\linewidth]{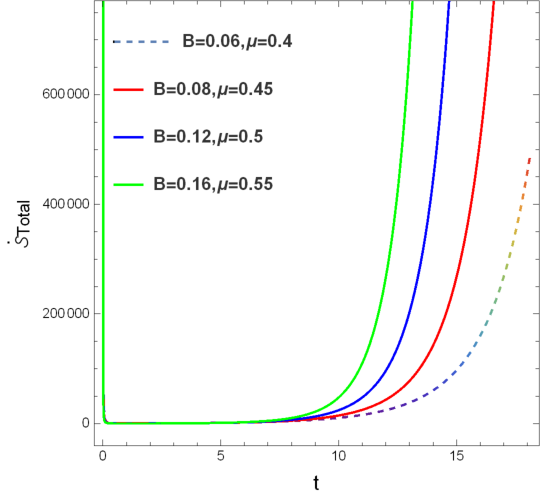}
\caption{Plot of $\dot{\mathcal{S}}$ to $t$ with time expressed in seconds, using Bekenstein entropy as entropy at the future event horizon with the background fluid as reconstructed viscous coupled Tsallis entropic holographic dark fluid.}
\label{Fig_12_TSallis_Sdot}
\end{minipage}%
\centering
\begin{minipage}{.5\textwidth}
  \centering
\includegraphics[width=.9\linewidth]{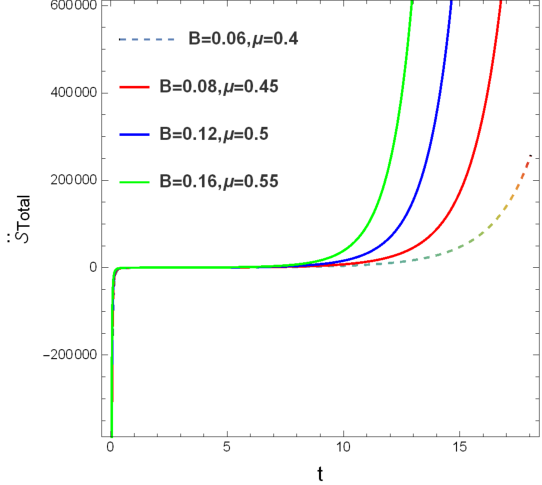}
\caption{Plot of thermal equilibrium $\ddot{\mathcal{S}}$ to $t$ with time expressed in seconds, using Bekenstein entropy as entropy at the future event horizon with the background fluid as reconstructed viscous coupled Tsallis entropic holographic dark fluid.}
\label{Fig_13_TSallis_Sddot}
\end{minipage}%
\end{figure}

This section examines the validity of GSL in terms of the future event horizon $R_E$. In order to identify the appropriate and well-behaved system's IR cut-off, a thorough examination of the Tsallis entropic HDE is required. We discuss the viscous-coupled Tsallis model as an example, considering the future event horizon, $R_E$. By giving particular horizons in cosmological models of an accelerated Universe an entropy, we can estimate the amount of information that lies behind them. Now, we confirm the validity of both GSLT and thermodynamic equilibrium for an isolated physical system with a maximum entropy state. The total of all constituent entropies (dark matter and DE, in this case) plus the entropy of the universe's horizon (the future event horizon), which can never decrease, is the definition of the second law of thermodynamics in the context of a cosmological system. So, by substituting $R_E$ from Eq. (\ref{RE}), $\rho_{m}$ from Eq. (\ref{Tsallis darkenergy density}), $\rho_{Total}$ from Eq. (\ref{total density TSallis}), and the coupled Tsallis-infused thermodynamic pressure $P$ and viscous pressure ($\Pi$) in Eqs. (\ref{entropy-fluid}), (\ref{entropy-horizon}), we can obtain the time derivative of total entropy in viscous interacting Tsallis HDE scenario. Since it meets the inequality $\dot{\mathcal{S}}\ge 0$, Fig\ref{Fig_12_TSallis_Sdot}. In addition, the thermal equilibrium (TE) for $\ddot{\mathcal{S}} \leq 0$ is established in Fig\ref{Fig_13_TSallis_Sddot}.
\subsection{Generalized Second Law of Thermodynamics at Particle Horizon}
\begin{figure}
\centering
\begin{minipage}{.5\textwidth}
  \centering
\includegraphics[width=.9\linewidth]{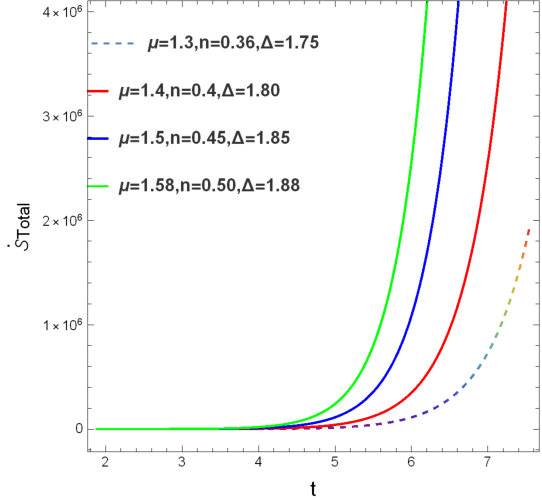}
\caption{Plot of $\dot{\mathcal{S}}$ to $t$ with time expressed in seconds, using Bekenstein entropy as entropy at the particle horizon with the background fluid as reconstructed viscous coupled Barrow entropic holographic dark fluid.}
\label{Fig_14_Barrow_Sdot}
\end{minipage}%
\centering
\begin{minipage}{.5\textwidth}
  \centering
\includegraphics[width=.9\linewidth]{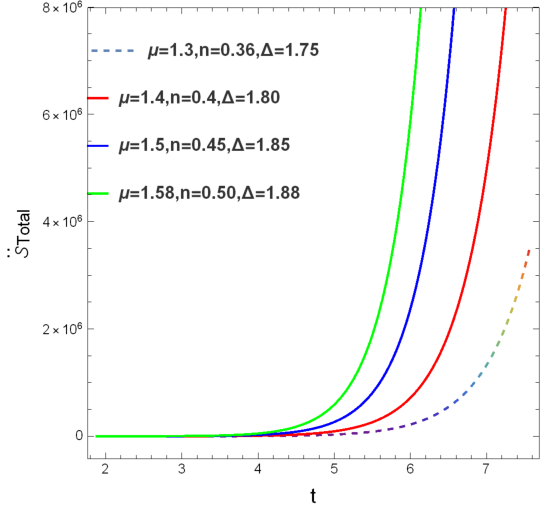}
\caption{Plot of thermal equilibrium $\ddot{\mathcal{S}}$ to $t$ with time expressed in seconds, using Bekenstein entropy as entropy at the particle horizon with the background fluid as reconstructed viscous coupled Barrow entropic holographic dark fluid.}
\label{Fig_15_Barrow_Sddot}
\end{minipage}%
\end{figure}

In this section we assume for the moment that there is an particle horizon in the FRW universe, which is a null space with vanishing expansion. The temperature and radius for the spatially flat FRW metric are specified as Eq. (\ref{RP}) with the background fluid as viscous coupled Barrow entropic HDE. So, by substituting $R_P$ from Eq. (\ref{RP}), $\rho_{m}$ from Eq. (\ref{BHDE darkmatter density}), $\rho_{Total}$ from Eq. (\ref{BHDE total density}), and the coupled Barrow HDE-infused thermodynamic pressure $P$ and viscous pressure ($\Pi$) in Eqs. (\ref{entropy-fluid}), (\ref{entropy-horizon}), we can obtain the time derivative of total entropy in viscous interacting Barrow HDE scenario. Since it meets the inequality $\dot{\mathcal{S}}\ge 0$, Fig\ref{Fig_14_Barrow_Sdot}.
\section{Generalized second law of thermodynamics under corrected entropies}
In this section, we examine the applicability of the GSL of thermodynamics for the different forms of the HDE model, where the entropy-area relation is modified at all the apparent event and particle horizons. As we previously discussed, we have considered two different sorts of modifications: the power-law correction and the logarithmic correction. This exercise aims to investigate the effects of these modifications on the GSL of thermodynamics for this specific interacting viscous HDE models.
\subsection{Thermodynamical Analysis for Logarithmic Corrected Entropy}
\begin{figure}[htb]
\centering 
\begin{subfigure}{0.25\textwidth}
\includegraphics[width=.7\linewidth]{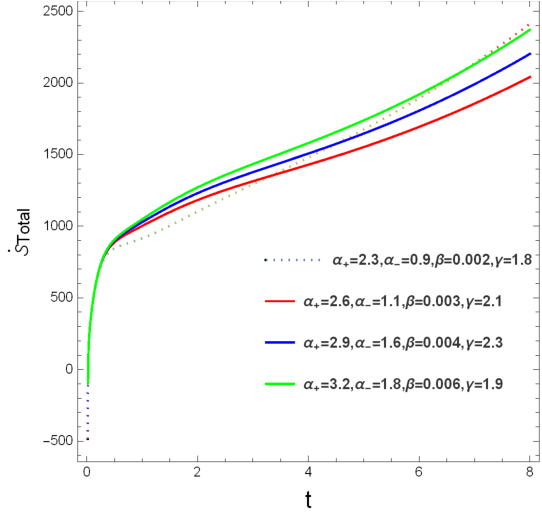}
\caption{Plot of $\dot{\mathcal{S}}$ to $t$ with time expressed in seconds, using logarithimic corrected entropy as entropy at the Hubble horizon with the background fluid as viscous coupled GHDE.}
\label{Fig_06_GHDE_logarithimic_Sdot}
\end{subfigure}\hfil 
\begin{subfigure}{0.25\textwidth}
 \includegraphics[width=.7\linewidth]{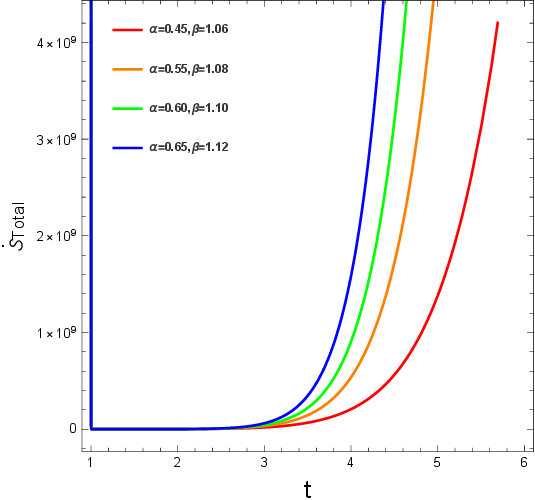}
  \caption{Plot of $\dot{\mathcal{S}}$ to $t$ with time expressed in seconds, using logarithimic corrected entropy as entropy at the Hubble horizon with the background fluid as viscous coupled NOHDE.}
  \label{Fig_16_NOHDE_Sdot_Logarithimic}
\end{subfigure}\hfil 
\begin{subfigure}{0.25\textwidth}
 \includegraphics[width=.7\linewidth]{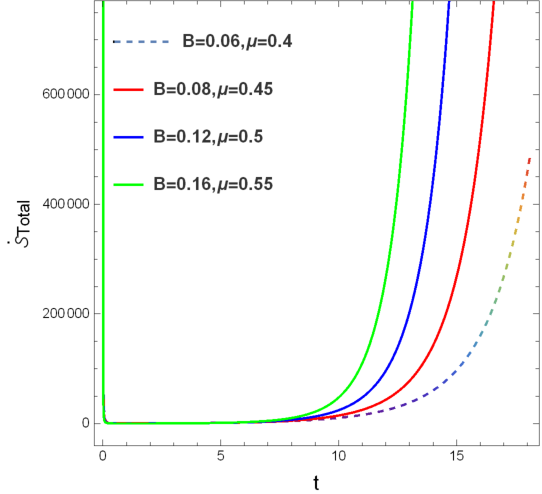}
  \caption{Plot of $\dot{\mathcal{S}}$ to $t$ with time expressed in seconds, using logarithimic corrected entropy as entropy at the future event horizon with the background fluid as viscous coupled THDE.}
  \label{Fig_20_TSallis_Sdot_Logarithimic}
\end{subfigure}\hfil 
\begin{subfigure}{0.25\textwidth}
  \includegraphics[width=.7\linewidth]{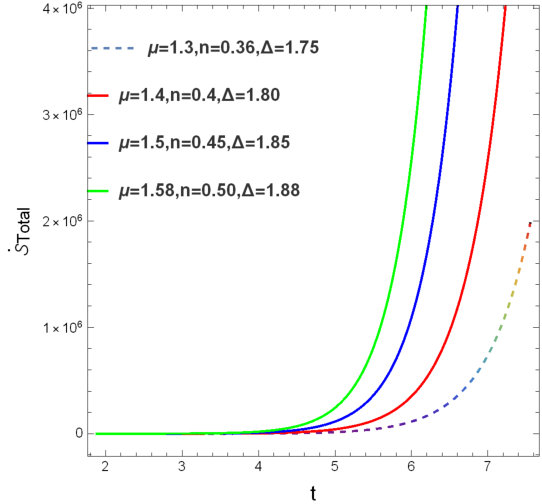}
  \caption{Plot of $\dot{\mathcal{S}}$ to $t$ with time expressed in seconds, using logarithimic corrected entropy as entropy at the particle horizon with the background fluid as viscous coupled BHDE.}
  \label{Fig_24_Barrow_Sdot_Logarithimic}
\end{subfigure}

\medskip
\begin{subfigure}{0.25\textwidth}
  \includegraphics[width=.7\linewidth]{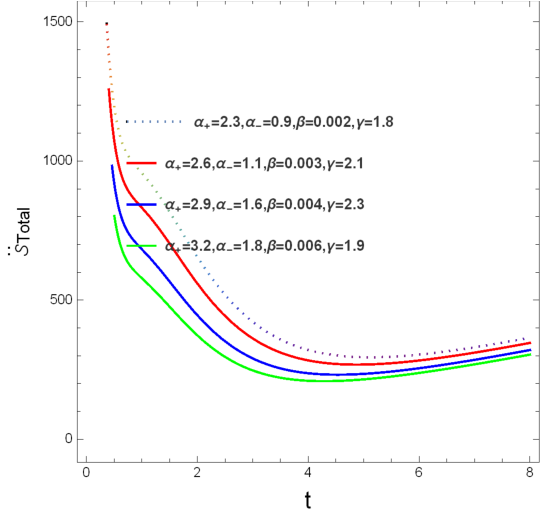}
  \caption{Plot of thermal equilibrium $\ddot{\mathcal{S}}$ to $t$ with time expressed in seconds, using logarithimic corrected entropy as entropy at the Hubble horizon with the background fluid as viscous coupled GHDE.}
  \label{Fig_07_GHDE_logarithimic_Sddot}
\end{subfigure}\hfil 
\begin{subfigure}{0.25\textwidth}
  \includegraphics[width=.7\linewidth]{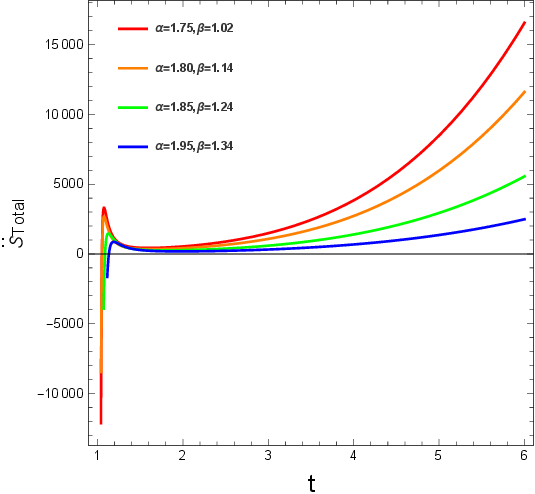}
  \caption{Plot of thermal equilibrium $\ddot{\mathcal{S}}$ to $t$ with time expressed in seconds, using logarithimic corrected entropy as entropy at the Hubble horizon with the background fluid as viscous coupled NOHDE.}
  \label{Fig_17_NOHDE_Sddot_Logarithimic}
\end{subfigure}\hfil 
\begin{subfigure}{0.25\textwidth}
  \includegraphics[width=.7\linewidth]{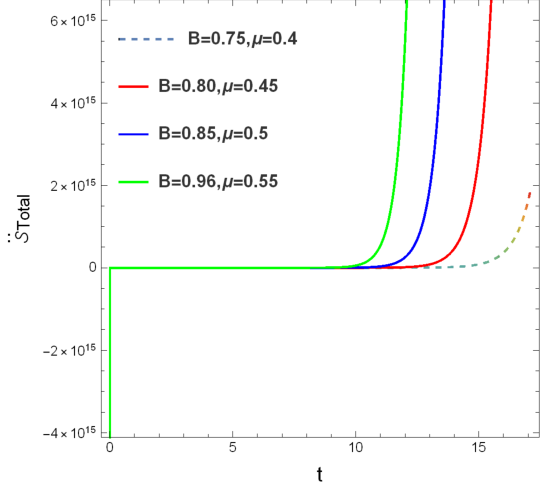}
  \caption{Plot of thermal equilibrium $\ddot{\mathcal{S}}$ to $t$ with time expressed in seconds, using logarithimic corrected entropy as entropy at the future event horizon with the background fluid as viscous coupled THDE.}
  \label{Fig_21_TSallis_Sddot_Logarithimic}
\end{subfigure}\hfil 
\begin{subfigure}{0.25\textwidth}
 \includegraphics[width=.7\linewidth]{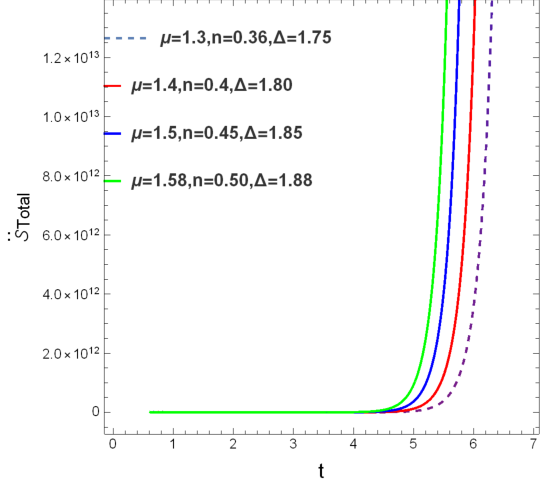}
  \caption{Plot of thermal equilibrium $\ddot{\mathcal{S}}$ to $t$ with time expressed in seconds, using logarithimic corrected entropy as entropy at the particle horizon with the background fluid as viscous coupled BHDE.}
  \label{Fig_25_Barrow_Sddot_Logarithimic}
\end{subfigure}
\caption{Logarithimic Entropy Correction}
\label{Logarithimic Entropy Correction}
\end{figure}

We address the horizon-related entropy addition in order to examine the universe's entropy growth. When both quantum and thermal equilibrium fluctuations are present, quantum gravity permits logarithmic corrections \cite{thermodynamic24,thermodynamic25,logarithimic1,logarithimic2,logarithimic3,logarithimic4,logarithimic5}. The corrected Wald entropy of horizons \cite{logarithimic5} can be obtained using quantum gravity as:
\begin{equation}
\mathcal{S} = \mathcal{S}_w + \alpha \ln{\mathcal{S}_w}+...,
\label{logarithimic correction2}
\end{equation}
The definition of logarithimic correction to entropy on perceived horizons is Eq. (\ref{logarithimic correction2}), which can also be obtained by modifying the entropy-area relation Eq. (\ref{Bekenstein-area}) and by examining the impact of the thermal fluctuations on the horizon entropy. Thus, we express the logarithmic entropy adjusted as follows,
\begin{equation}
\mathcal{S}_h=\frac{A}{4G}+\upsilon \ln{\frac{A}{4G}}+\Psi \frac{4 G}{A}+\psi 
\label{logarithimic horizon}
\end{equation}
The logarithimic entropy in the context of 4-parameter entropic most generalized GHDE can be found by substituting $R_H$ from Eq. (\ref{R-H}), $\rho_{m}$ from Eq. (\ref{4-parameter dark-matter density}), $\rho_{Total}$ from Eq. (\ref{total density 4-parameter}), and thermodynamic pressure $P$ from Eq. (\ref{pressure 4-parameter}) and viscous pressure ($\Pi$) infused by coupled GHDE in Eqs. (\ref{entropy-fluid}), (\ref{logarithimic horizon}). This allows us to obtain the time derivative of total logarithimic corrected entropy in viscous coupled 4-parameter entropic reconstruction based most generalized GHDE scenario. In the context of NOHDE, the logarithimic entropy can be found by replacing $R_H$ from Eq. (\ref{RH}), $\rho_{m}$ from Eq. (\ref{density-darkmatter-NOHDE}), $\rho_{Total}$ from Eq. (\ref{total density-NOHDE}) and thermodynamic pressure $P$ and viscous pressure ($\Pi$) infused by coupled NOHDE in Eqs.(\ref{entropy-fluid}), (\ref{logarithimic horizon}), we can get the time derivative of total logarithimic corrected entropy in viscous coupled NOHDE scenario. The total of all constituent entropies (dark matter and DE, in this case) plus the entropy of the universe's horizon (the future event horizon), which can never decrease, is the definition of the second law of thermodynamics in the context of a cosmological system. So, for finding time derivative of total logarithimic entropy in the context of viscous coupled Tsallis entropic HDE, we can substitute $R_E$ from Eq. (\ref{RE}), $\rho_{m}$ from Eq. (\ref{Tsallis darkenergy density}), $\rho_{Total}$ from Eq. (\ref{total density TSallis}), and the coupled Tsallis-infused thermodynamic pressure $P$ and viscous pressure ($\Pi$) in Eqs. (\ref{entropy-fluid}), (\ref{logarithimic horizon}), and hence, we can obtain the time derivative of total logarithimic corrected entropy in viscous interacting Tsallis HDE scenario. Again,  So, by substituting $R_P$ from Eq. (\ref{RP}), $\rho_{m}$ from Eq. (\ref{BHDE darkmatter density}), $\rho_{Total}$ from Eq. (\ref{BHDE total density}), and the coupled Barrow HDE-infused thermodynamic pressure $P$ and viscous pressure ($\Pi$) in Eqs. (\ref{entropy-fluid}), (\ref{logarithimic horizon}), we can obtain the time derivative of total logarithimic corrected entropy in viscous interacting Barrow HDE scenario. Since, logarithimic corrected entropy meets the inequality $\dot{\mathcal{S}}\ge 0$ for all the cases of HDE, in Figs\ref{Fig_06_GHDE_logarithimic_Sdot},\ref{Fig_16_NOHDE_Sdot_Logarithimic}, \ref{Fig_20_TSallis_Sdot_Logarithimic} and\ref{Fig_24_Barrow_Sdot_Logarithimic}. The thermal equilibrium inequality $\dot{\mathcal{S}}\ge 0$, meets for all the forms of HDE i.e GHDE, NOHDE, Tsallis and Barrow in Figs\ref{Fig_07_GHDE_logarithimic_Sddot},\ref{Fig_17_NOHDE_Sddot_Logarithimic}, \ref{Fig_21_TSallis_Sddot_Logarithimic} and \ref{Fig_25_Barrow_Sddot_Logarithimic} respectively.

\subsection{Thermodynamical Analysis for Power-Law Corrected Entropy}
\begin{figure}[htb]
    \centering 
\begin{subfigure}{0.25\textwidth}
\includegraphics[width=.7\linewidth]{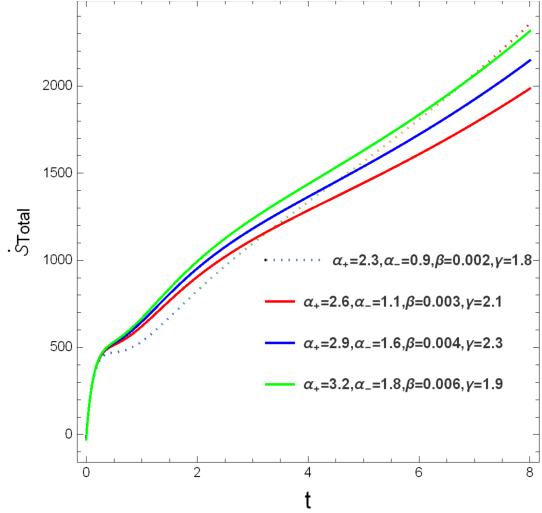}
\caption{Plot of $\dot{\mathcal{S}}$ to $t$ with time expressed in seconds, using power-law corrected entropy as entropy at the Hubble horizon with the background fluid as viscous coupled GHDE.}
\label{Fig_08_Powerlaw_GHDE_Sdot}
\end{subfigure}\hfil 
\begin{subfigure}{0.25\textwidth}
 \includegraphics[width=.7\linewidth]{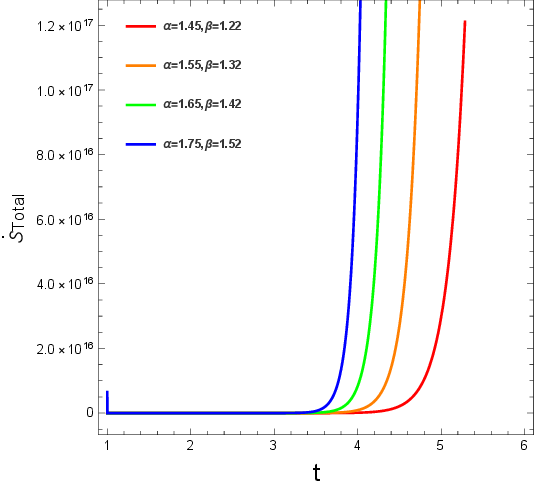}
  \caption{Plot of $\dot{\mathcal{S}}$ to $t$ with time expressed in seconds, using power-law corrected entropy as entropy at the Hubble horizon with the background fluid as viscous coupled NOHDE.}
  \label{Fig_18_NOHDE_Sdot_Powerlaw}
\end{subfigure}\hfil 
\begin{subfigure}{0.25\textwidth}
 \includegraphics[width=.7\linewidth]{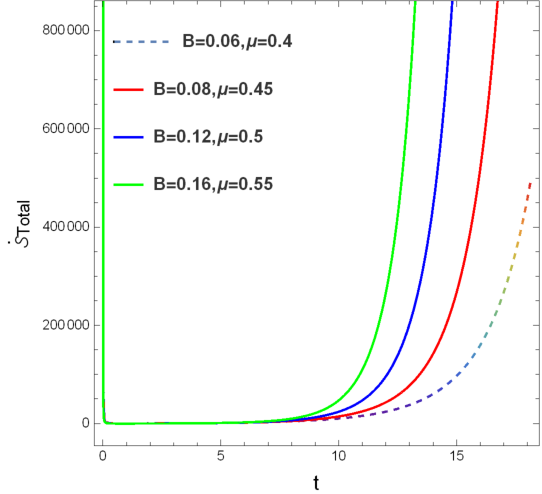}
  \caption{Plot of $\dot{\mathcal{S}}$ to $t$ with time expressed in seconds, using power-law corrected entropy as entropy at the future event horizon with the background fluid as viscous coupled THDE.}
  \label{Fig_22_TSallis_Sdot_Powerlaw}
\end{subfigure}\hfil 
\begin{subfigure}{0.25\textwidth}
  \includegraphics[width=.7\linewidth]{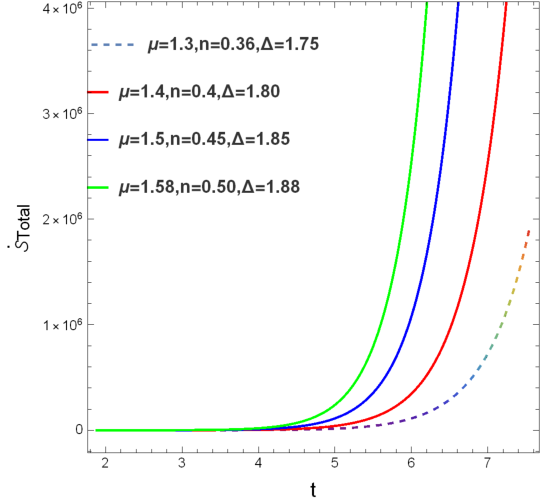}
  \caption{Plot of $\dot{\mathcal{S}}$ to $t$ with time expressed in seconds, using power-law corrected entropy as entropy at the particle horizon with the background fluid as viscous coupled BHDE.}
  \label{Fig_26_Barrow_Sdot_Powerlaw}
\end{subfigure}

\medskip
\begin{subfigure}{0.25\textwidth}
  \includegraphics[width=.7\linewidth]{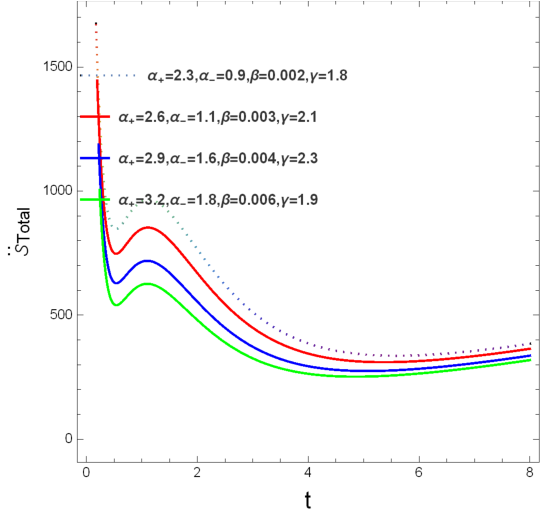}
  \caption{Plot of thermal equilibrium $\ddot{\mathcal{S}}$ to $t$ with time expressed in seconds, using power-law corrected entropy as entropy at the Hubble horizon with the background fluid as viscous coupled GHDE.}
  \label{Fig_09_Powerlaw_GHDE_Sddot}
\end{subfigure}\hfil 
\begin{subfigure}{0.25\textwidth}
  \includegraphics[width=.7\linewidth]{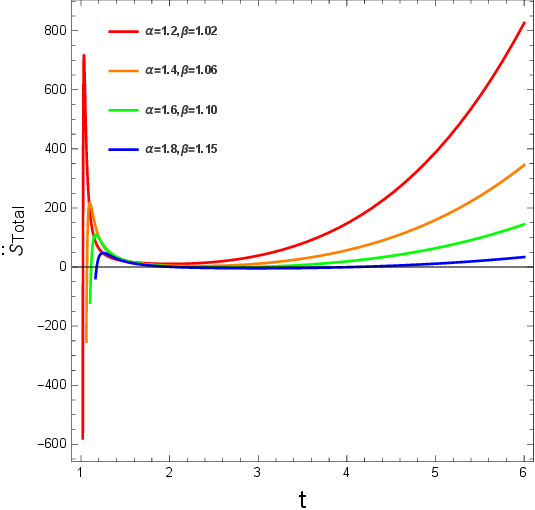}
  \caption{Plot of thermal equilibrium $\ddot{\mathcal{S}}$ to $t$ with time expressed in seconds, using power-law corrected entropy as entropy at the Hubble horizon with the background fluid as viscous coupled NOHDE.}
  \label{Fig_19_NOHDE_Sddot_Powerlaw}
\end{subfigure}\hfil 
\begin{subfigure}{0.25\textwidth}
  \includegraphics[width=.7\linewidth]{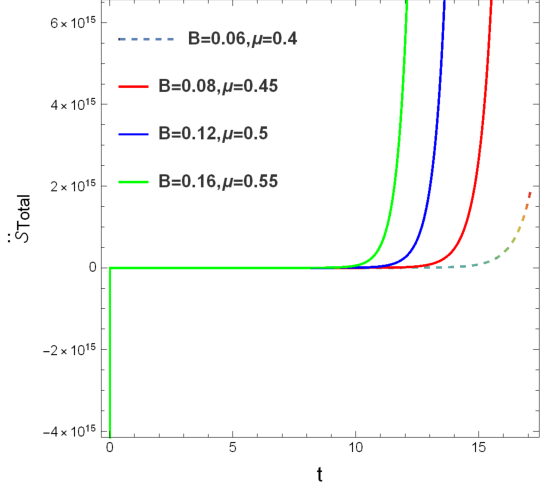}
  \caption{Plot of thermal equilibrium $\ddot{\mathcal{S}}$ to $t$ with time expressed in seconds, using power-law corrected entropy as entropy at the future event horizon with the background fluid as viscous coupled THDE.}
  \label{Fig_23_TSallis_Sddot_Powerlaw}
\end{subfigure}\hfil 
\begin{subfigure}{0.25\textwidth}
 \includegraphics[width=.7\linewidth]{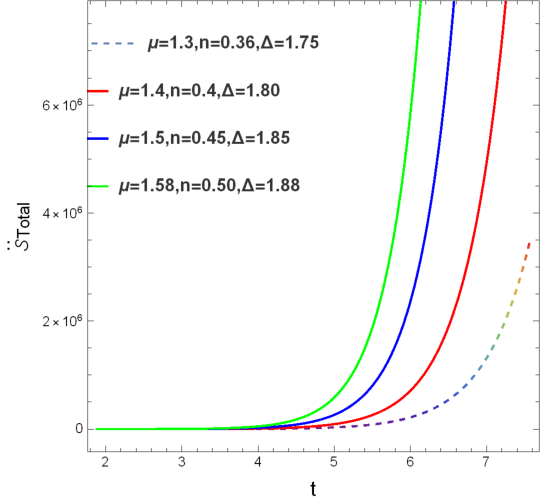}
  \caption{Plot of thermal equilibrium $\ddot{\mathcal{S}}$ to $t$ with time expressed in seconds, using power-law corrected entropy as entropy at the particle horizon with the background fluid as viscous coupled BHDE.}
  \label{Fig_27_Barrow_Sddot_Powerlaw}
\end{subfigure}
\caption{Power-Law Entropy Correction}
\label{Power-Law Entropy Correction}
\end{figure}

In the context of quantum field entanglement in and out of the horizon, power law corrections to entropy are observed \cite{Power-Law1}.The modified power-law form of entropy\cite{Power-Law2} is expressed as:
\begin{equation}
\mathcal{S}_h=\frac{A}{4} [1-K_\upsilon A^{1-\frac{\upsilon}{2}}],
\label{powerlaw horizon}
\end{equation}
where, $A=4\pi R_h^2$ and \textbf{$K_\upsilon=\frac{\upsilon(4\pi)^{\frac{\upsilon}{2}-1}}{(4-\upsilon)r_c^{2-\upsilon}}$}. In the context of 4-parameter entropic GHDE, the power-law entropy can be found by replacing $R_H$ from Eq. (\ref{R-H}), $\rho_{m}$ from Eq. (\ref{4-parameter dark-matter density}), $\rho_{Total}$ from Eq. (\ref{total density 4-parameter}), and in Eqs. (\ref{entropy-fluid}), (\ref{powerlaw horizon}), thermodynamic pressure $P$ from Eq. (\ref{pressure 4-parameter}) and viscous pressure ($\Pi$) infused by coupled GHDE. Again, the power-law entropy with the background fluid as viscous NOHDE can be found by substituting $R_H$ from Eq. (\ref{RH}), $\rho_{m}$ from Eq. (\ref{density-darkmatter-NOHDE}), $\rho_{Total}$ from Eq. (\ref{total density-NOHDE}), and thermodynamic pressure $P$ and viscous pressure ($\Pi$) infused by coupled NOHDE in Eqs.(\ref{entropy-fluid}), (\ref{powerlaw horizon}). This allows us to obtain the time derivative of total power-law corrected entropy in viscous coupled NOHDE scenario. The definition of the second law of thermodynamics in the context of a cosmological system is the sum of all constituent entropies (dark matter and DE in this example) plus the entropy of the universe's horizon (the future event horizon), which can never decrease. Thus, in the case of viscous coupled Tsallis entropic HDE, we can substitute $R_E$ from Eq. (\ref{RE}), $\rho_{m}$ from Eq. (\ref{Tsallis darkenergy density}), $\rho_{Total}$ from Eq. (\ref{total density TSallis}), and the coupled Tsallis-infused thermodynamic pressure $P$ and viscous pressure ($\Pi$) in Eqs. (\ref{entropy-fluid}) and (\ref{powerlaw horizon}) to find the time derivative of total power-law entropy.We can thus determine the time derivative of total power-law corrected entropy in the viscous interacting Tsallis HDE scenario by taking into account. Therefore, we can obtain the time derivative of total power-law corrected entropy in viscous interacting Barrow HDE scenario by substituting $R_P$ from Eq. (\ref{RP}), $\rho_{m}$ from Eq. (\ref{BHDE darkmatter density}), $\rho_{Total}$ from Eq. (\ref{BHDE total density}), and the coupled Barrow HDE-infused thermodynamic pressure $P$ and viscous pressure ($\Pi$) in Eqs. (\ref{entropy-fluid}), (\ref{powerlaw horizon}). As a result, in Figs\ref{Fig_08_Powerlaw_GHDE_Sdot},\ref{Fig_18_NOHDE_Sdot_Powerlaw}, \ref{Fig_22_TSallis_Sdot_Powerlaw}, and \ref{Fig_26_Barrow_Sdot_Powerlaw}, power-law corrected entropy satisfies the inequality $\dot{\mathcal{S}}\ge 0$ for all viscous coupled HDE situations. In Figs\ref{Fig_09_Powerlaw_GHDE_Sddot},\ref{Fig_19_NOHDE_Sddot_Powerlaw}, \ref{Fig_23_TSallis_Sddot_Powerlaw}, and \ref{Fig_26_Barrow_Sdot_Powerlaw}, respectively, the thermal equilibrium inequality $\dot{\mathcal{S}}\ge 0$ is satisfied for all kinds of HDE, namely 4-parametric GHDE, NOHDE, Tsallis, and Barrow in power-law corrected entropy case.

\section{Conclusion}
Black hole physics provides the foundation for the concept of thermodynamics in cosmic systems. The temperature of Hawking radiations emitted from black holes is hypothesized to be proportional to their corresponding surface gravity on the event horizon \cite{entropy1}. Jacobson \cite{entropy2} discovered a connection between Einstein's field equations and thermodynamics. He arrived at this relationship using the entropy-horizon area proportionality relation and the first law of thermodynamics, also known as the Clausius relation: $dQ = \mathcal{T} d\mathcal{S}$, where $dQ$, $\mathcal{T}$, $d\mathcal{S}$ and represent the energy, temperature, and entropy change exchanged for a particular system. The field equations for any spherically symmetric spacetime were demonstrated to be written as follows: $\mathcal{T} d\mathcal{S}= dE + P dV$ for any horizon (where $E$, $P$, and $V$ stand for the internal energy, pressure, and volume of the spherical system) \cite{entropy3}. This relationship has allowed GSLT to be thoroughly explored in the context of the universe's expanding behavior. The universe's horizon entropy, which is used to discuss GSLT, can be expressed as one-fourth of its horizon area \cite{entropy4}, power law corrected \cite{entropy5,entropy6,entropy7}, or logarithmic corrected \cite{entropy8}. Many researchers have used the universe's simple horizon entropy to investigate the validity of GSLT in a variety of systems, such as the interaction of two fluid components, such as DE and dark matter \cite{entropy9,entropy10,entropy11,entropy12}, as well as the interaction of three fluid components, such as \cite{entropy13,entropy14,entropy15}, in the FRW world. Much work was done on modified theories of gravity using thermodynamic analysis \cite{entropy16,entropy17,entropy18,entropy19}.

In this paper, we proposed a new four-parameter generalized entropy function based viscous coupled holographic dark fluid reconstruction with appropriate parameter limitations. This novel four-parameter entropy function ($\mathcal{S}_g$) that extends the capabilities of the Tsallis, Renyi, Barrow, Sharma Mittal, Kaniadakis, and Loop Quantum Gravity entropies. We investigate here whether the reconstructed holographic energy density corresponding to the 4-parameter generalized entropy can be the only factor driving the evolution of the early universe, specifically from inflation, based on the viscous coupled entropic GHDE reconstruction scenario. The least complex form of the generalized version of entropy that may reduce to all known entropies that have been proposed to date for appropriate limitations of the entropic parameters is the 4-parameter generalized entropy (shown by $\mathcal{S}_g$).
Inspired by the works as mentioned earlier, we have examined a recently proposed 4-parameter entropic generalized HDE model which demonstrates different entropic holographic dark energy in the flat FRW world. Initially, we developed four cosmological holographic dark energy models by selecting four IR cut-offs: the most generalized 4-parameter entropic holographic dark energy, the generalized holographic dark energy with the Nojiri-Odintsov cut-off, the Tsallis entropic holographic dark fluid, and the Barrow entropic dark fluid in a viscous interacting setting. In the case of GHDE and NOHDE, we selected the Hubble horizon as the maximum length of the cut-off, and also included the logamediate scale factor for further calculations. 

The NOHDE,Tsallis, and Barrow HDE are particular cases of the most generalized form of 4-parameter entropic generalized HDE(GHDE) proposed by Nojiri-Odintsov as already mentioned. Without presuming any particular relationship between the pressure and energy density of the Universe, we have explored the role of a hybrid scale factor in achieving feasible cosmic dynamics in this current study. We have considered 4-parameter entropic generalized holographic reconstruction based on apparent horizon with hybrid scale factor. Between the power law and exponential law expansion behaviors, there is an intermediary, the hybrid scale factor. The hybrid scale factor consists of two components: the first dominates early in the evolution of the universe and behaves like the power law expansion. At late phases, the other factor takes over and exhibits exponential growth behavior, resulting in a model that is more like to the concordance $\Lambda$CDM model. The viscous interacting EoS parameter in the 4-parameter entropic most generalized viscous coupled GHDE scenario is shown in Fig\ref{Fig_01_GHDE_w}, whereas the effective EoS parameter of the reconstructed equivalent entropic dark energy scenario is shown in Fig\ref{Fig_02_GHDE_weff}. The EoS parameter displays phantom behavior in Fig\ref{Fig_01_GHDE_w} and is asymptotically close to the phantom boundary in late time, whereas the effective EoS parameter in Fig\ref{Fig_2_NO_weff} is rapidly approaching the phantom boundary without overcoming the Big-rip singularity.
Here, we can see our model is stable under small perturbation in Fig\ref{Fig_02_GHDE_weff} in an interacting most generalized coupled GHDE scenario. In this study, the effective EoS parameter of the reconstructed NOHDE is shown in Fig\ref{Fig_2_NO_weff}, whereas the viscous interacting EoS parameter is shown in Fig\ref{Fig_1_NO_w} in the logamediate case. While the effective EoS parameter in Fig\ref{Fig_2_NO_weff} is able to overcome the Big-rip singularity, it is gradually approaching the phantom boundary in Fig\ref{Fig_1_NO_w}, where it displays its quintessential behavior and is crossing the phantom boundary in late time. Similarly, we have chosen the future event and particle horizons for Tsallis entropic HDE and Barrow entropic HDE, respectively. The EoS parameter $w$ expression in the interacting viscous scenario is derived and plotted against time in Fig{\ref{Fig_4_Tsallis_w}}. By using the intermediate scale factor, the expressions of $\rho_{effective}$ and $\Pi$ in Eq. (\ref{2nd-Fried}), we can obtain the expression of pressure $(P)$. We have demonstrated $w_{eff}=\frac{P+\pi}{\rho_{Total}}$ in an interacting viscous reconstructed THDE scenario in Fig\ref{Fig_5_Tsallis_weff}. The interacting EoS parameter $w$ in Fig\ref{Fig_4_Tsallis_w} is seen to cross the $-1$ phantom barrier early in the time series. After crossing the $-1$ phantom barrier, the interacting EoS parameter $w$ can exhibit quintessential behavior and prevent big-rip singularity in the later phase period. In Fig\ref{Fig_5_Tsallis_weff}, the interacting effective EoS parameter serves as a quintessence. In the emergent viscous interacting scenario, the effective EoS parameter is shown in Fig{\ref{Fig_8_BHDE_weff}}, and the expression of the EoS parameter $w$ in the interacting viscous BHDE scenario is determined and plotted against time in Fig\ref{Fig_7_BHDE_w}. We can observe that the model is stable under small perturbations since $c_s^2 \ge 0$ after examining its stability in this viscosity interaction BHDE scenario. It is observed that the interacting EoS parameter $w$ in Fig\ref{Fig_7_BHDE_w} exhibits characteristic behavior. The interacting effective EoS parameter in Fig\ref{Fig_8_BHDE_weff} acts as ghost behavior in later phases of the universe without preventing Big-rip singularity. Figs\ref{Fig_3_Cs-square}, \ref{Fig_6_Tsallis_Cs-square}, \ref{Fig_9_BHDE_Cs-square} illustrate that all of our models is stable under small perturbations in the viscosity-interacting HDE scenario. Now, we have created thermodynamical quantities and examined the applicability of thermodynamic equilibrium and GSLT. We have made a number of entropy correction assumptions for the rich development of the current DE model's thermodynamics, including Bekenstein, logarithmic correction, and power law entropies at the universe's apparent, future event, and particle horizons. We discovered that GSLT holds true for all entropy and horizon scenarios. Additionally, we met the thermal equilibrium requirement for specific constant parameter conditions. These are the specific outcomes in detail.

In the apparent horizon, with the background fluid as viscous coupled GHDE and NOHDE, we have explored GSLT using standard Bekenstein entropy. In the framework of a cosmological system, the second law of thermodynamics is defined as the total entropy of all the components (mostly dark matter and DE) plus the constant entropy of the universe's border (the Hubble horizon in this case). Therefore, by substituting $R_H$ from Eq. (\ref{R-H}), $\rho_{m}$ from Eq. (\ref{4-parameter dark-matter density}), $\rho_{Total}$ from Eq. (\ref{total density 4-parameter}), thermodynamic pressure $P$ from Eq. (\ref{pressure 4-parameter}), and viscous pressure ($\Pi$) infused by coupled GHDE in Eqs. (\ref{entropy-fluid}), (\ref{entropy-horizon}), we can obtain the time derivative of total entropy in a viscous interacting 4-parameter entropic GHDE scenario. Figure\ref{Fig_04_GHDE_Sdot} plots it. The time derivative of total entropy is consistent with Fig\ref{Fig_04_GHDE_Sdot} since it satisfies the inequality $\dot{\mathcal{S}}\ge 0$. Furthermore, $\ddot{\mathcal{S}} \leq 0$ has its thermal equilibrium (TE) established in late-time of evolution of the Universe although at the early phases TE is not established.  Fig\ref{Fig_10_NOHDE_Sdot}, which illustrates GSLT on the apparent horizon using standard Bekenstein entropy, validates the validity of GSLT by demonstrating that $\dot{\mathcal{S}}\ge 0$ with the increase of time $t$. Additionally, Fig\ref{Fig_11_NOHDE_Sddot} shows that the thermal equilibrium criterion for Bekenstein entropy at the apparent horizon is satisfied. Figs\ref{Fig_06_GHDE_logarithimic_Sdot} and \ref{Fig_16_NOHDE_Sdot_Logarithimic} show the GSLT on apparent horizon for logarithmic corrected entropy and show that GSLT is still valid for all values of $t$ with the background fluid as GHDE and NOHDE respectively. Consequently, the thermal equilibrium requirement is satisfied for all logarithmic entropy at apparent horizon. Figs\ref{Fig_08_Powerlaw_GHDE_Sdot} and \ref{Fig_18_NOHDE_Sdot_Powerlaw} show the plots of $\dot{\mathcal{S}}$ by taking power-law corrected entropy at apparent horizon, which show the validity of GSLT and acts favorably with time with the background fluid as GHDE and NOHDE. In addition, with this entropy, the power-law corrected entropy at apparent horizon criterion for thermal equilibrium has been met ( Fig\ref{Fig_19_NOHDE_Sddot_Powerlaw}). As illustrated in the aforementioned figure, $t$ is plotted at apparent horizon by using power law entropy at apparent horizon. In this case, increasing $t$ with positive moves certifies the efficacy of GSLT at apparent horizon. 

Regarding the event horizon, with the background fluid as viscous-coupled THDE, Fig\ref{Fig_12_TSallis_Sdot} shows that at the event horizon with Bekenstein entropy, GSLT is still applicable. In addition, the Bekenstein entropy met the requirements for thermodynamic equilibrium at the event horizon. Fig\ref{Fig_13_TSallis_Sddot}. When logarithmic entropy is present, the validity of GSLT is confirmed at the event horizon by Fig\ref{Fig_20_TSallis_Sdot_Logarithimic}. Fig\ref{Fig_21_TSallis_Sddot_Logarithimic} shows that for $t > 0$, supporting the validity of the thermal equilibrium condition. Fig\ref{Fig_22_TSallis_Sdot_Powerlaw} demonstrates that when $t$ increases, the trajectories of $\dot{\mathcal{S}}$ stay positive. This indicates that the current situation complies with the requirements for thermodynamic equilibrium. Fig\ref{Fig_23_TSallis_Sddot_Powerlaw} shows the plot of $\ddot{\mathcal{S}}$ to $t$ for power law corrected entropy. Note that in this case, GSLT holds. Fig\ref{Fig_23_TSallis_Sddot_Powerlaw} illustrates how the current situation satisfies the requirements of thermodynamic equilibrium for power-law entropy at the event horizon.

With reference to the particle horizon, with the background fluid as viscous-coupled BHDE, GSLT is still valid at the event horizon with Bekenstein entropy, as shown in Fig\ref{Fig_14_Barrow_Sdot}. Furthermore, the Bekenstein entropy at the particle horizon satisfied the conditions for thermodynamic equilibrium. Illustration \ref{Fig_15_Barrow_Sddot}. Fig\ref{Fig_24_Barrow_Sdot_Logarithimic} validates the validity of GSLT at the particle horizon when logarithmic entropy is present. For $t > 0$, as shown in Fig\ref{Fig_25_Barrow_Sddot_Logarithimic}, the thermal equilibrium condition is valid. The trajectories of $\dot{\mathcal{S}}$ remain positive as $t$ grows, as shown in Fig\ref{Fig_26_Barrow_Sdot_Powerlaw}. This suggests that the state of affairs now meets the conditions for thermodynamic equilibrium. The power law corrected entropy plot of $\ddot{\mathcal{S}}$ to $t$ is presented in Fig\ref{Fig_27_Barrow_Sddot_Powerlaw}. Note that in this case, GSLT holds. Fig\ref{Fig_27_Barrow_Sddot_Powerlaw} illustrates how the current situation satisfies the requirements of thermodynamic equilibrium for power-law entropy at the event horizon.

\section{Declaration of competing interest}
The authors state that none of their known competing financial interests or personal connections could have impacted the work that came out in this paper.

\section{acknowledgement}
Sanghati Saha acknowledges the hospitality of the Inter-University Centre for Astronomy and Astrophysics (IUCAA), Pune, India, for their hospitality during her scientific visit from August 25 to September 6, 2023. Surajit Chattopadhyay acknowledges the visiting associateship of IUCAA.

\end{document}